\documentclass[%
 aip,
 amsmath,amssymb,
 reprint,%
]{revtex4-1}

\usepackage{graphicx}
\usepackage{dcolumn}
\usepackage{bm}

\usepackage[utf8]{inputenc}
\usepackage[T1]{fontenc}
\usepackage{mathptmx}
\usepackage{etoolbox}

\makeatletter
\def\@email#1#2{%
 \endgroup
 \patchcmd{\titleblock@produce}
  {\frontmatter@RRAPformat}
  {\frontmatter@RRAPformat{\produce@RRAP{*#1\href{mailto:#2}{#2}}}\frontmatter@RRAPformat}
  {}{}
}%
\makeatother
\begin{document}

\preprint{AIP/123-QED}

\title[]{Effective Phonon Dispersion and Low field transport in Al$_x$Ga$_{1-x}$N alloys using supercells: An ab-initio approach}
\author{Animesh Datta}
\author{Ankit Sharma}
\author{Matinehsadat Hosseinigheidari}
\author{Uttam Singisetti}%

 \email{uttamsin@buffalo.edu}

\date{\today}

\begin{abstract}
To investigate the transport properties in random alloys, it is important to model the alloy disorder using supercells. Though traditional methods like Virtual Crystal Approximation (VCA) are computationally efficient, the local disorder in the system is not accurately captured as artificial translational symmetry is imposed on the system. However, in the case of supercells, the error introduced by self-image interaction between the impurities is reduced and translational symmetry is explicitly imposed over larger length scales. In this work, we have investigated the Effective Phonon Dispersion (EPD) and transport properties, from first principle calculations using supercells in Al$_x$Ga$_{1-x}$N alloy systems. Using our in-house developed code, the EPD of AlGaN is obtained and the individual modes are identified. Next, we discuss our in-house developed method to calculate low-field transport properties in supercells. First to validate our methods we have solved the Boltzmann Transport Equation using Rode’s method to compare the phonon limited mobility in the 4 atom GaN primitive cell and 12 atom GaN supercell. Using the same technique, we have investigated the low field transport in random Al$_x$Ga$_{1-x}$N  alloy systems. Our calculations show that along with alloy scattering, electron-phonon scattering may also play an important role at room temperature and high-temperature device operation. This technique opens up the path for calculating phonon-limited transport properties in random alloy systems.
\end{abstract}

\maketitle

\section{\label{sec:level1}Introduction and Overview}
In the last decade, ultra-wide bandgap (UWBG) materials have gained significant interest for power electronics applications. Several materials such as SiC\cite{casady1996status}, GaN\cite{nakamura2013blue,morkocc1999general,vurgaftman2003band,khan1991vertical}, Ga$_2$O$_3$\cite{higashiwaki2014development,wang2018band} has been explored extensively for RF, power, ultraviolet(UV) LEDs and space applications due to their promising properties. Among these materials, there has been significant success of nitride materials for light emitting diodes (LEDs), ultraviolet (UV) LEDs \cite{khan2008ultraviolet,pimputkar2009prospects} and high electron mobility transistors (HEMTs)\cite{mishra2002algan,mishra2008gan}. By alloying AlN with GaN, Al$_x$Ga$_{1-x}$N alloys provide more scope for device engineering due to the band gap tunability ranging from 3.4eV to 6.2eV\cite{strite1992gan}. However, there are still challenges associated with AlGaN-based devices such as energy efficiency. To realize the full potential of AlGaN-based devices, it is important to understand the processes that affect the transport properties of these materials. 
\\
Conventionally, Virtual Crystal Approximation (VCA) has been used to study the electronic properties of alloys, such as band gap and effective masses. However, VCA-based approximations fail to capture the true disorder of the random alloy system by imposing artificial translational symmetry. Thus recently, Kyrtsos et.al. \cite{kyrtsos2019first} performed a first-principles study of the electronic properties of AlGaN alloys using supercells. Supercells help to capture the true disorder of the alloy system since the error introduced by self-image interaction is reduced and translational symmetry is explicitly imposed over large length scales. In regards to transport properties, theoretical studies using analytical models\cite{coltrin2017transport} show that alloy disorder scattering is one of the most dominant scattering mechanisms which limits its electrical efficiency in AlGaN-based devices. The effects of alloy disorder scattering have also been extensively studied using first-principles methods using supercells\cite{pant2020high} as well as using corrections to the  Virtual Crystal Approximation (VCA) techniques\cite{bellotti2007alloy}. Previously using Monte Carlo simulations on semi-empirical band structures, there have also been reports of phonon and ionized impurity scattering limited mobility in AlGaN alloys\cite{farahmand2001monte}. However, there has been no reports of effective phonon dispersion and first principles study of phonon-limited transport properties in AlGaN alloys. It is important to understand the effect of the electron-phonon scattering process which is important at room temperature and higher temperatures in addition to alloy scattering.
Understanding the thermal properties of the semiconductor is critical for high-temperature device operation which can be studied by modeling the phonon dispersion.
\\
 Thus, in this work, we have investigated the Effective Phonon Dispersion(EPD) of AlGaN using supercells (SC) instead of virtual crystal interpolation techniques, and then based on first principles-based calculations we have developed a method to investigate the low-field phonon limited mobility in Al$_x$Ga$_{1-x}$N alloys. Our calculations are based on the parameters obtained from ab initio DFPT calculations of supercells. Firstly, using our in-house developed code explored in our companion paper, we try to understand the Effective Phonon Dispersion of Al$_x$Ga$_{1-x}$N for different Al fractions and identify the different phonon modes. However, it becomes challenging to calculate the transport properties from our unfolded dispersion due to numerically induced noise in the calculations. Thus we use our in-house developed method to calculate the low field phonon limited mobility using SC. First, we have verified our SC-based method using GaN primitive cells and SC. Finally, the technique is applied to AlGaN alloy systems to understand the effect of electron-phonon scattering on low-field mobility for different Al fraction concentrations.
\section{Computational Details}
To capture the full disorder of the alloy system, SQS (Special Quasirandom Structures)\cite{zunger1990special,wei1990electronic,wang1998majority} are generated using the Alloy Theoretic Automated Toolkit (ATAT)\cite{van2013efficient,van2009multicomponent}. For calculating the unfolded phonon dispersion of Al$_x$Ga$_{1-x}$N alloy system, a 40-atom SC is generated from the 10-atom PC as the fundamental repeating unit in the $5\times2\times1$ configuration. The generated supercell was subjected to volume and atomic relaxation such that the forces on each atom are less than $1\times10^{-4}Ry/au$. Subsequently the self-consistent calculations were performed using Quantum Espresso\cite{giannozzi2009quantum,giannozzi2017advanced,gonze1997dynamical,gonze1995adiabatic} on a $\Gamma$ centered $4\times4\times4$ reciprocal space $\vec{\textbf{k}}$-point grid with a planewave energy cutoff of 80Ry (1088eV). Then DFPT was used to calculate the phonon on a reciprocal space $\Gamma$ centered $4\times4\times4$ $\vec{\textbf{q}}$-point grid. Then using the in-house developed unfolding code, the EPD is obtained. 
\\
 The long-range electron-phonon interaction matrix elements contributing to polar optical phonon and piezoelectric scattering are calculated on a a 30$\times$30$\times$30 q grid using the method described by Verdi and Giustino\cite{verdi2015frohlich}, which is a generalization of the Frohlich interaction as implemented in EPW\cite{ponce2016epw}.  Using the calculated matrix elements, the polar optical phonon scattering and the piezoelectric scattering rate is calculated using Fermi's Golden Rule with an energy cut-off of 0.35eV from the conduction band minimum. The ionized impurity scattering is modeled using the Brooks-Herring model\cite{chattopadhyay1981electron}. We have also assumed partial ionization of dopants with an ionization energy of 27meV\cite{roccaforte2022ion} for GaN and AlGaN alloys assuming Si as potential shallow donor. For AlN, the dopant ionization energy was assumed to be 294meV \cite{kanechika2006n} according to previous experimental data. The deformation potential scattering rate is included using analytical equations\cite{}, since it doesn't play a significant role in low-field transport calculations. The effective mass of GaN is taken to be 0.22m$_e$\cite{mavroidis2003detailed,huang2001hall} and for AlN it is taken as 0.4m$_e$\cite{xu1993electronic} as reported in previous theoretical and experimental studies. In the case of alloys, the alloy disorder scattering is modeled using a statistically averaged disorder potential in the matrix element for Fermi's Golden Rule\cite{bellotti2007alloy,pant2020high}. The effective mass of intermediate AlGaN alloys is estimated using Vegard's Law\cite{dreyer2013effects}. Then Rode's iterative method is used to solve the Boltzmann Transport Equations to calculate the electron distribution and the low field mobility on a  30$\times$30$\times$30 q grid and 60$\times$60$\times$60 k grid. To account for the anisotropy of the EPI elements, reciprocal space integration is carried out numerically using the technique of Gaussian smearing of 5 meV for the energy conservation $\delta$ function
\section{Effective Phonon Dispersion of AlGaN}
For a 40-atom supercell conventional DFPT calculations obtained by Fourier interpolation of the dynamical matrix give 120 phonon modes. The large number of phonon modes becomes clustered within the energy window making it difficult to distinctly identify the individual modes. Thus we have used our in-house developed program to calculate the Effective Phonon Dispersion (EPD) which folds the 120 modes into 12 modes. The EPD formalism is based on the scheme proposed by Boykin\cite{boykin2014brillouin} to obtain the projection probability as calculated in the case of electronic band structure unfolding. 
\\
The orthonormality of the phonon eigenstates is used 
 to calculate the projection probability as given in  Eq.\ref{eq: u7} \cite{boykin2014brillouin}
\begin{equation}
    P(E_p,\vec{q}_m) = \sum_{\alpha,w}|C_p^{(\alpha,w)}|^2
    \label{eq: u7}
\end{equation}

\begin{figure}[h]
\begin{center}
    \includegraphics[width=1\linewidth]{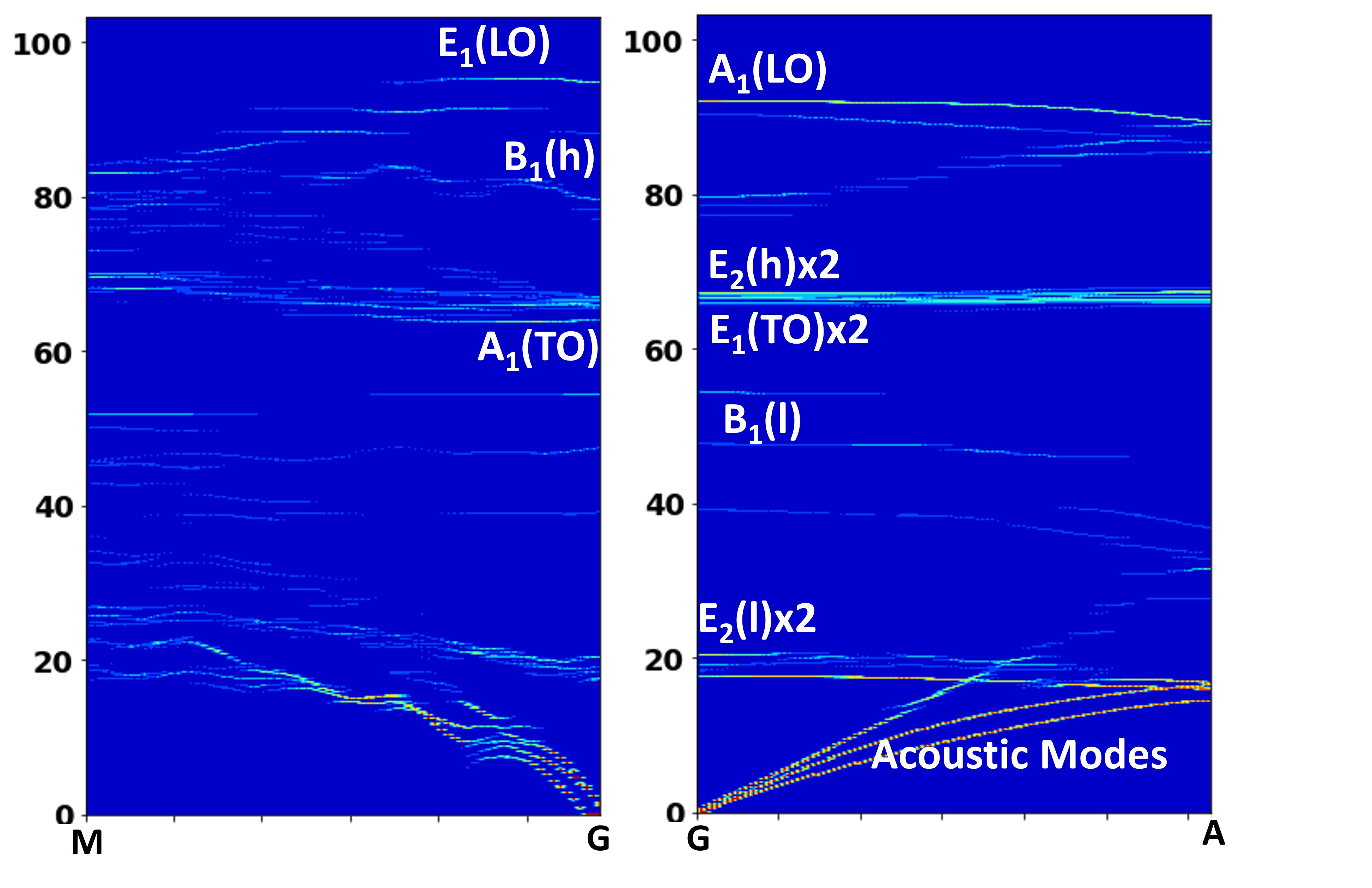}
    \caption{Effective Phonon Dispersion for 40 atom 25$\%$ Al fraction in the M-G-A direction}
    \label{epd25}
    \end{center}
\end{figure}

\begin{figure}[h]
\begin{center}
    \includegraphics[width=1\linewidth]{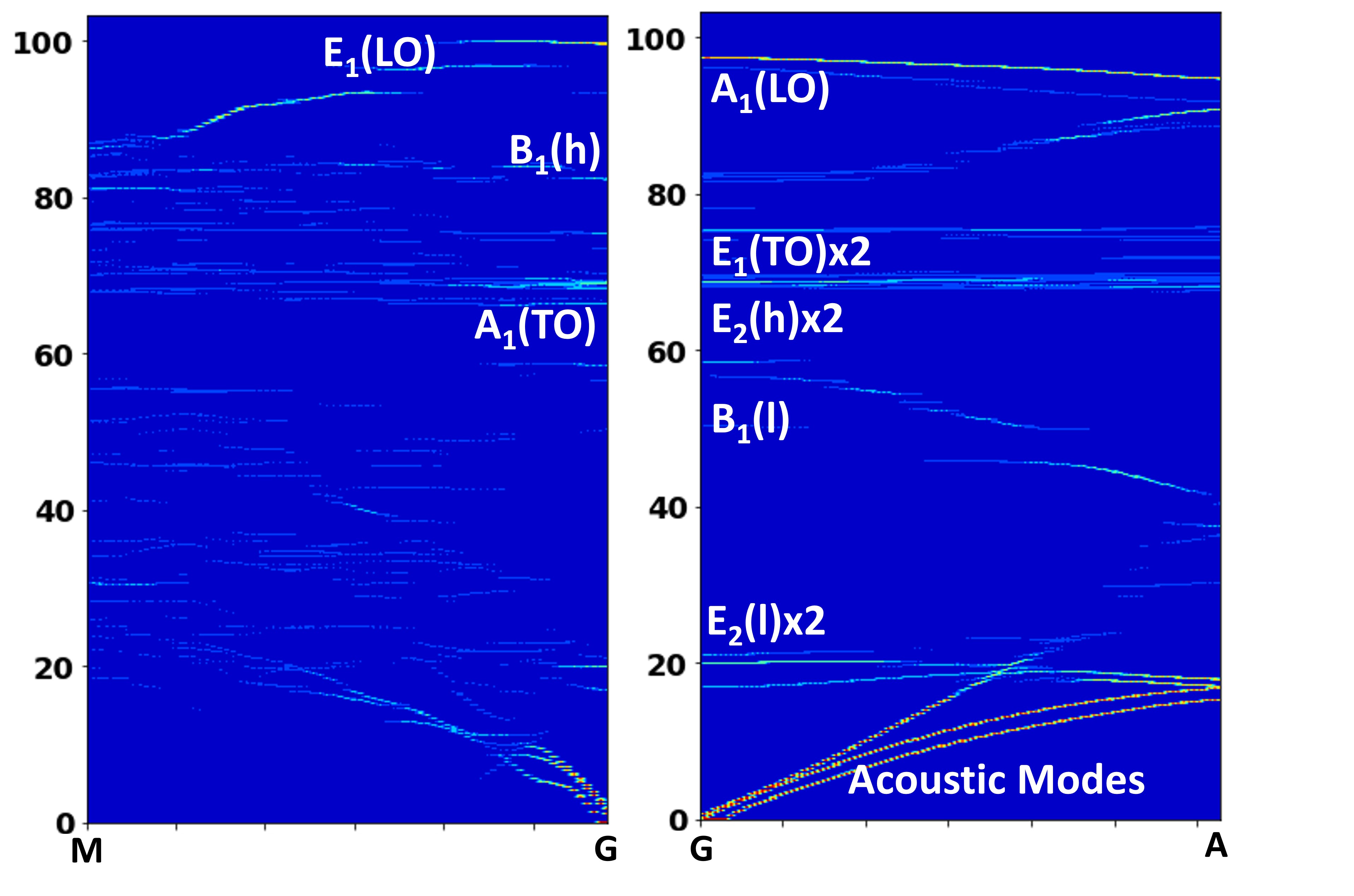}
    \caption{Effective Phonon Dispersion for 40 atom 50$\%$ Al fraction in the M-G-A direction}
    \label{epd50}
    \end{center}
\end{figure}

The projection probability is used to calculate the phonon spectral function which eventually gives us the Effective Phonon Dispersion. The expression for the spectral function is given in Eq.\ref{eq: u9}
\begin{equation}
    A(\epsilon,\vec{q})_m) = \sum_{p=1}^{3rNc}P(E_p,\vec{q}_m)\delta(\epsilon-E_p)
    \label{eq: u9}
\end{equation}
where \textbf{A} is the spectral function, \textbf{p} runs over all the SC modes, $E_p$ is the supercell phonon eigenenergy in mode \textbf{p} and wavevector $\vec{Q}$ and $\epsilon$ is the energy grid with an increment of 0.01eV. The $\delta$ function is approximated as a Gaussian with a standard deviation of 0.0025eV.
\\
Fig.\ref{epd25} and \ref{epd50} shows the EPD of Al$_{0.25}$Ga$_{0.75}$N alloys and Al$_{0.5}$Ga$_{0.5}$N respectively in the M-G-A direction. 
It is observed from Fig.\ref{epd25} and \ref{epd50}, that the acoustic modes are stronger and the optical modes appear diffused which can be attributed to the alloy-induced disorder. The overall frequency range also increases with the increasing Al fraction which is in agreement with the experimentally and theoretically verified blue shift phenomenon\cite{holtz2001composition}. From the EPD curves, the individual modes have been identified distinctly. Our calculations for the E$_1$(LO), A$_1$(LO), E$_2$(h), A$_1$(TO) mode energies shows good match with experimentally observed Raman Spectra data for Al$_{x}$Ga$_{1-x}$N alloys\cite{holtz2001composition,davydov2002composition}.
\begin{table}[!ht]
    \centering
    \begin{tabular}{c|c|c}
    \hline
    \multicolumn{3}{c}{\textbf{Al$_{0.25}$Ga$_{0.75}$N}} \\
    \hline
    Mode & This Work & Experiment \\
    ~ & (\textbf{meV}) & (\textbf{meV})\\
    \hline
    \hline
        E$_1$(LO) &95 & 96.1 \cite{davydov2002composition} \\
        A$_1$(LO) & 93 & 95.48 \cite{davydov2002composition}, 94.24\cite{holtz2001composition} \\ 
        E$_1$(TO) & 66.1 & 70.6\cite{davydov2002composition} \\ 
        E$_2$(h) & 67.4 & 70.68\cite{davydov2002composition} \\ 
        A$_1$(TO) & 65 & 67.58\cite{davydov2002composition}, 65.72\cite{holtz2001composition} \\ \hline
    \end{tabular}
    \caption{
        Calculated and experimental phonon energies at Gamma Point for Al$_{0.25}$Ga$_{0.75}$N}
    \label{al25}
\end{table}

\begin{table}[!ht]
    \centering
    \begin{tabular}{c|c|c}
    \hline
    \multicolumn{3}{c}{\textbf{Al$_{0.5}$Ga$_{0.5}$N}} \\
    \hline
    Mode & This Work & Experiment \\
    ~ & (\textbf{meV}) & (\textbf{meV})\\
    \hline
    \hline
        E$_1$(LO) &99 & 100.75 \cite{davydov2002composition} \\
        A$_1$(LO) & 97 & 99.82 \cite{davydov2002composition}, 100.44\cite{holtz2001composition} \\ 
        E$_1$(TO) & 70 & 71.3\cite{davydov2002composition}\\
        E$_2$(h) & 69 & 71.17\cite{davydov2002composition}\\
        A$_1$(TO) & 68 & 69.192\cite{davydov2002composition}\\ \hline
    \end{tabular}
    \caption{Calculated and experimental phonon energies at Gamma Point for Al$_{0.5}$Ga$_{0.5}$N}
    \label{al50}
\end{table}
In the lower Al fraction range, the modes behave similar to GaN but as we further increase the Al fraction in the alloy, the modes tend to behave more like AlN-like modes. However, due to numerical noise in our simulations, the silent B$_1$(l) mode is diffused and couldn't be identified distinctly as the others. This technique opens up the path to identify the individual phonon modes in AlGaN alloys which play a crucial role in thermal conductivity properties.

\section{Phonon-limited mobility in supercells}
\begin{figure}[h]
\begin{center}
    \includegraphics[width=0.5\textwidth]{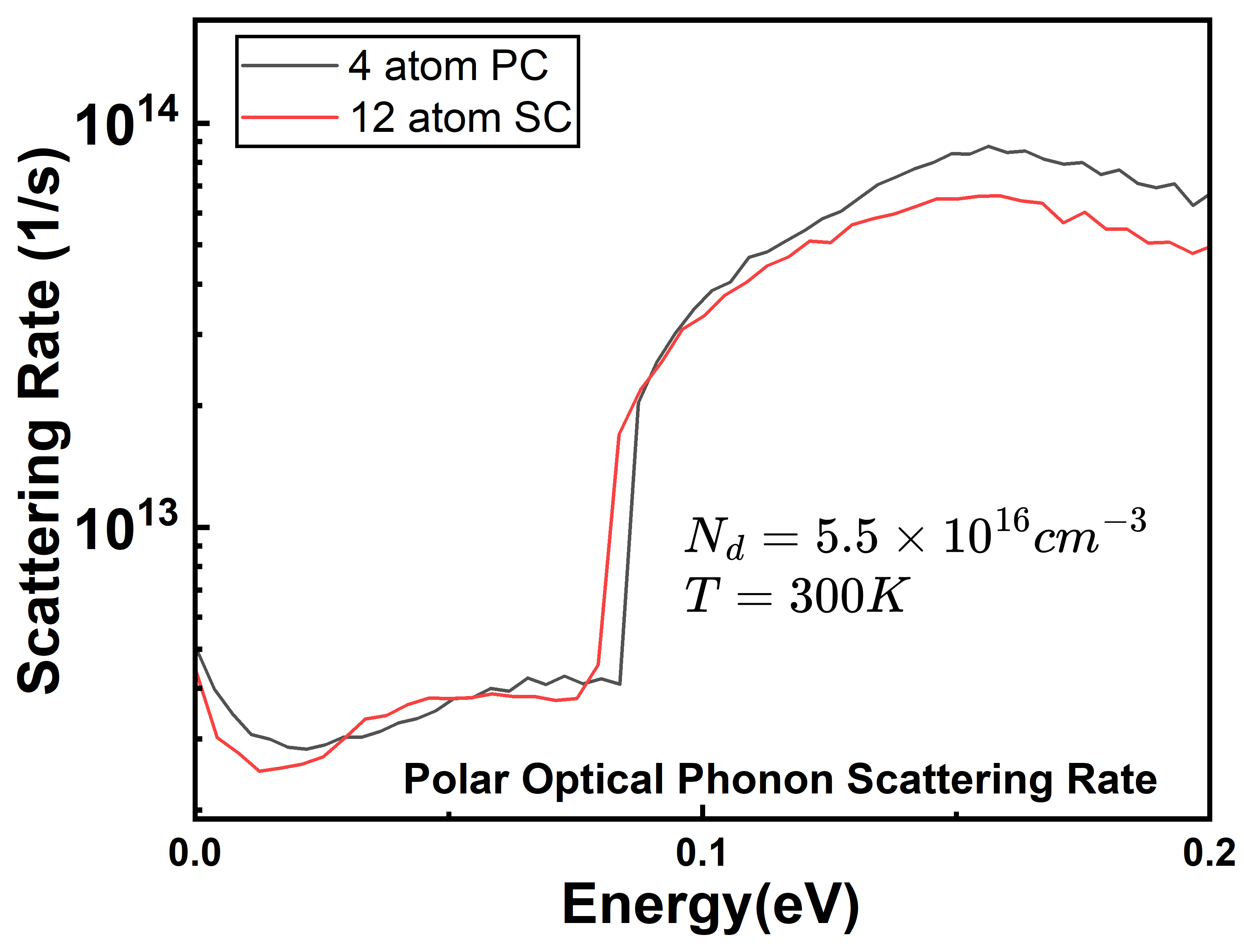}
    \caption{Polar Optical Phonon Scattering Rate for GaN PC and SC}
    \label{ganpcsc}
    \end{center}
\end{figure}

\begin{figure}[h]
\begin{center}
    \includegraphics[width=\linewidth]{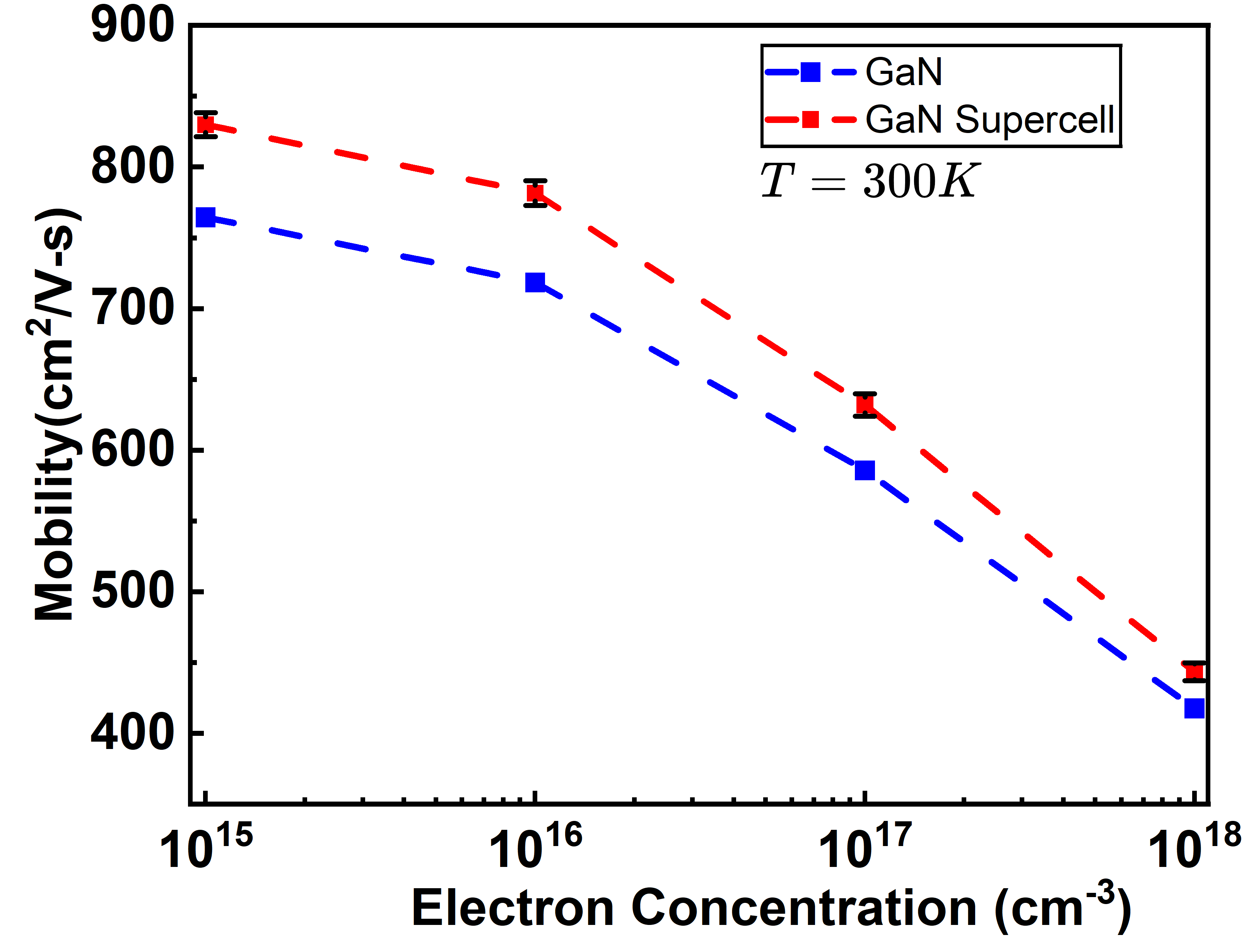}
    \caption{Calculated low field mobility at various electron concentrations for GaN PC and SC}
    \label{ganpcscmob}
    \end{center}
\end{figure}

 The EPD formalism gives us an understanding of the broadening of the phonon modes due to alloy disorder and the characteristics of the individual modes. However numerical noise in the EPDs makes it challenging to calculate the transport properties from the unfolded phonon spectra. It also becomes difficult to calculate the unfolded phonon displacement vectors from the SC to PC atoms which are required for evaluating the electron-phonon interaction matrix elements. Thus in this section, we discuss our in-house developed method to calculate the phonon-limited low-field mobility directly from supercells. In low-field transport calculations, the short-range deformation potentials do not play a significant part, thus in our calculations the acoustic deformation potential and optical deformation potential scattering mechanisms are modeled analytically. The long-range polar optical phonon scattering and the piezoelectric scattering mechanisms are calculated from ab initio first principle calculations. The ionized impurity scattering is modeled using the Brooks-Herring model.\cite{chattopadhyay1981electron}
\\ 
To validate our method, we verify our calculations on primitive GaN and 12-atom GaN supercell. From conventional DFPT calculation on GaN primitive cell, there are 3 IR active modes and ideally, it shouldn't change irrespective of the cell size. However, DFPT calculations on GaN supercell yield more than 3 IR active modes which in turn affects the polar optical phonon scattering rate. Thus to account for the more IR active modes, we have modified the q-grid range while calculating the long-range electron-phonon matrix elements. In the case of supercells, the q grid range is modified according to the volume ratio of the supercell and primitive cell. From the physics point of view, we know that the scattering strength of the long-range matrix elements decays as we move away from the gamma point further into the Brillouin Zone. Since supercells have more IR active modes, the modified q grid allows the scattering strength of all the modes to decay completely and thus comparable polar optical phonon scattering rates are obtained for primitive and supercell configurations. Fig.\ref{ganpcsc} shows the polar optical phonon scattering rate in both GaN primitive cell and 12 atom GaN supercell for an electron concentration of n=5.5e16 cm$^{-3}$ at T=300K. 
The scattering rates show a close match and thus serve as the first verification of our method. Then using Rode's iterative method we calculated the low field mobility of GaN at different electron concentrations at room temperature for primitive cells and supercell. Fig.\ref{ganpcscmob} shows the calculated low field mobility of GaN primitive cell and supercell for a range of electron concentrations at room temperature.
From Fig.\ref{ganpcscmob} it can be seen that the low field mobility values for both supercell and primitive cell show a close match with an error percentage of 8$\%$. Our calculated values for the low field GaN mobility is also comparable with previous ab initio-based studies and experimental values of mobility in GaN.\cite{ponce2019hole,kyle2014high,gotz1998hall} The mobility calculations serve as a validation for our method to calculate the low field mobility in SC. This method has enabled the calculation of low field mobility in disordered alloy systems through a first principles-based approach.
\\
In a recent study, Jhalani et.al\cite{jhalani2020piezoelectric} investigated the effect of dynamic quadrupoles on long-range electron-phonon interaction. It was seen that including the quadrupole interaction suppresses the contribution of acoustic phonons to long-range scattering mechanisms and thus the mobility increases significantly. Though in this section, we have not included this interaction in our calculations, we have discussed this effect in the next section while investigating the low-field transport in AlGaN alloys. 
\section{Low field transport in AlGaN alloys}
\begin{figure}
\begin{center}
    \includegraphics[width=\linewidth]{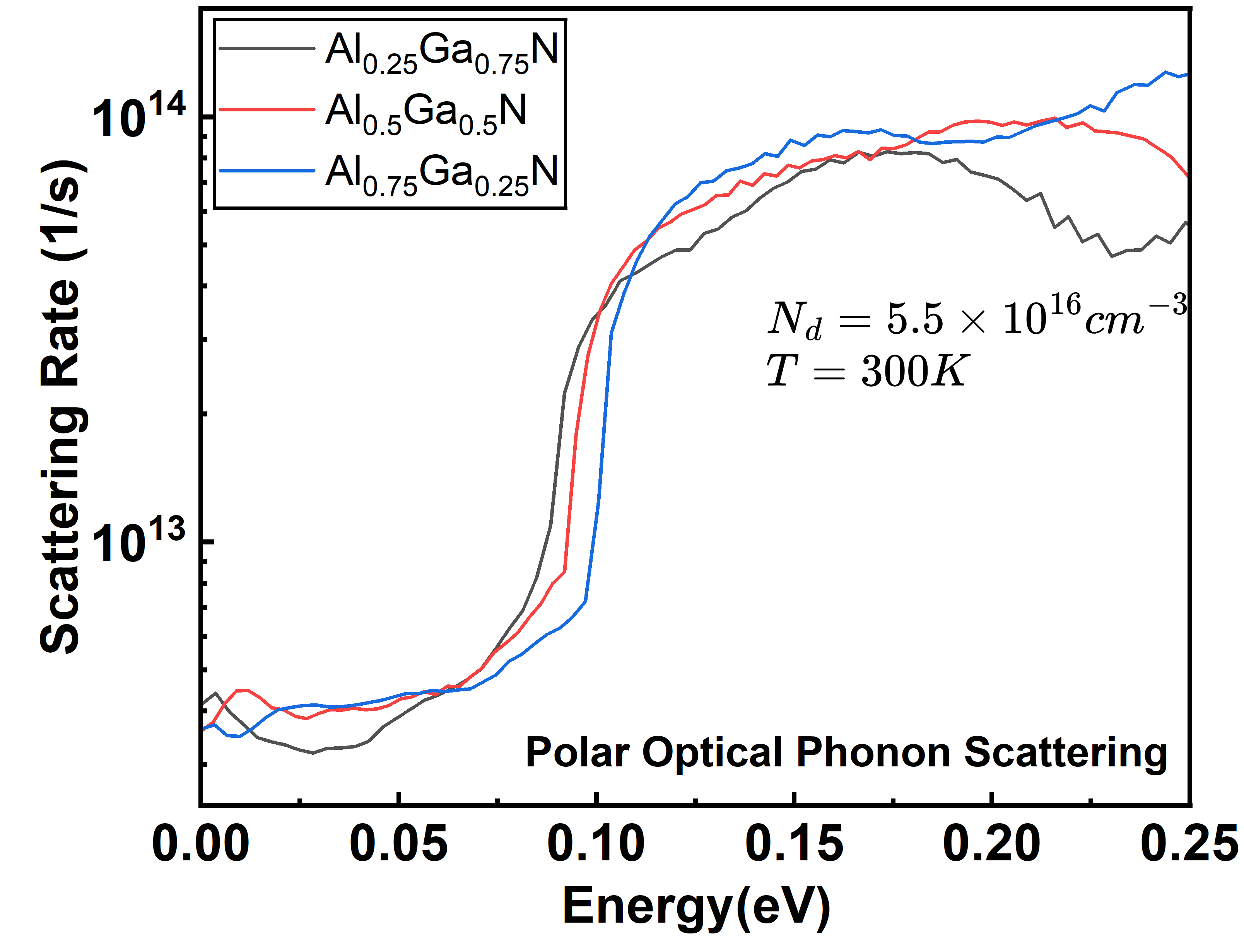}
    \caption{Polar Optical Phonon Scattering Rate for AlGaN at different Al concentrations}
    \label{alganscat}
    \end{center}
\end{figure}

\begin{figure}[h!]
    \begin{minipage}[c]{1\linewidth}
        \centering
        \includegraphics[width=1\textwidth]{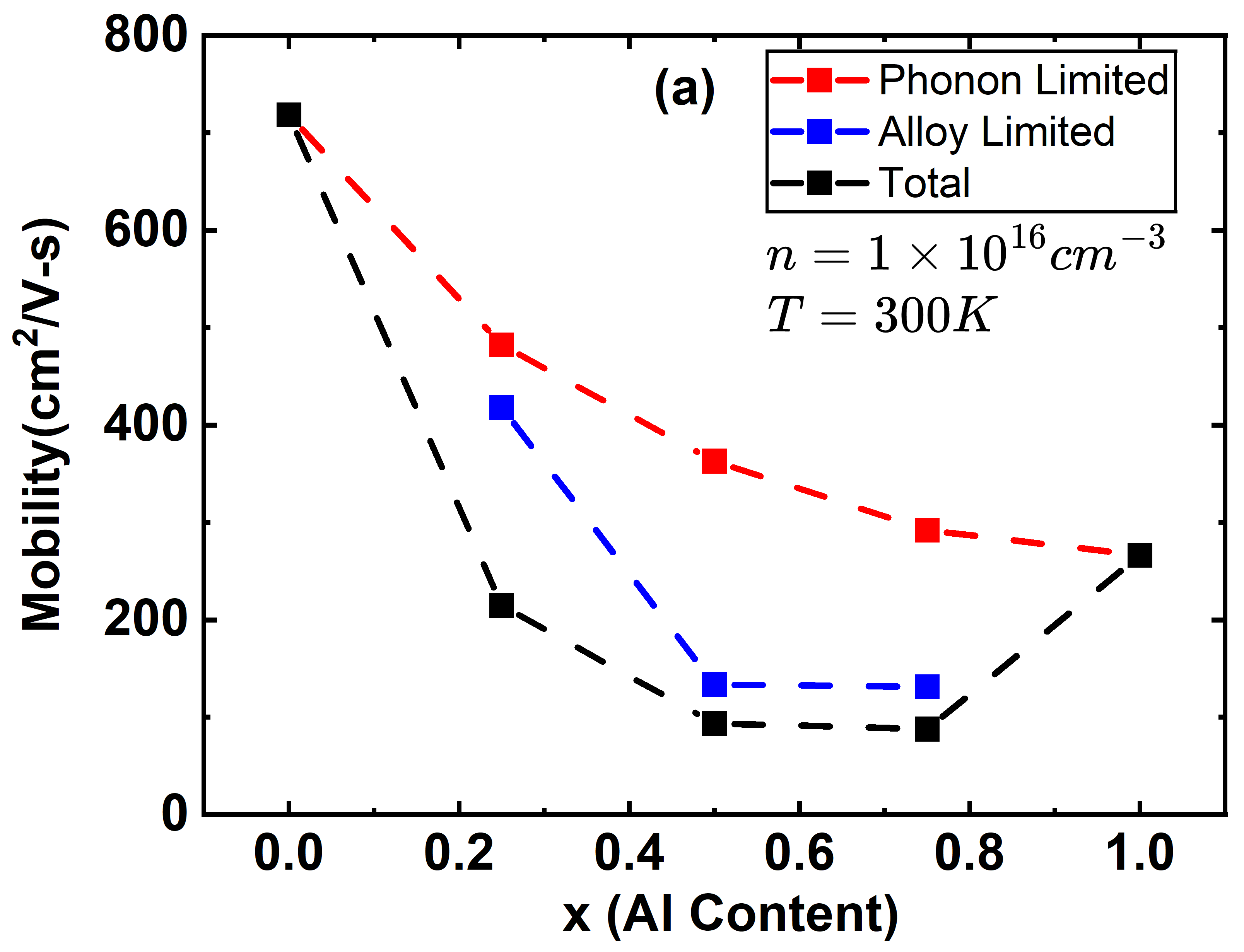}
    \end{minipage}\hfill
    \begin{minipage}[c]{1\linewidth}
        \centering
        \includegraphics[width=1\textwidth]{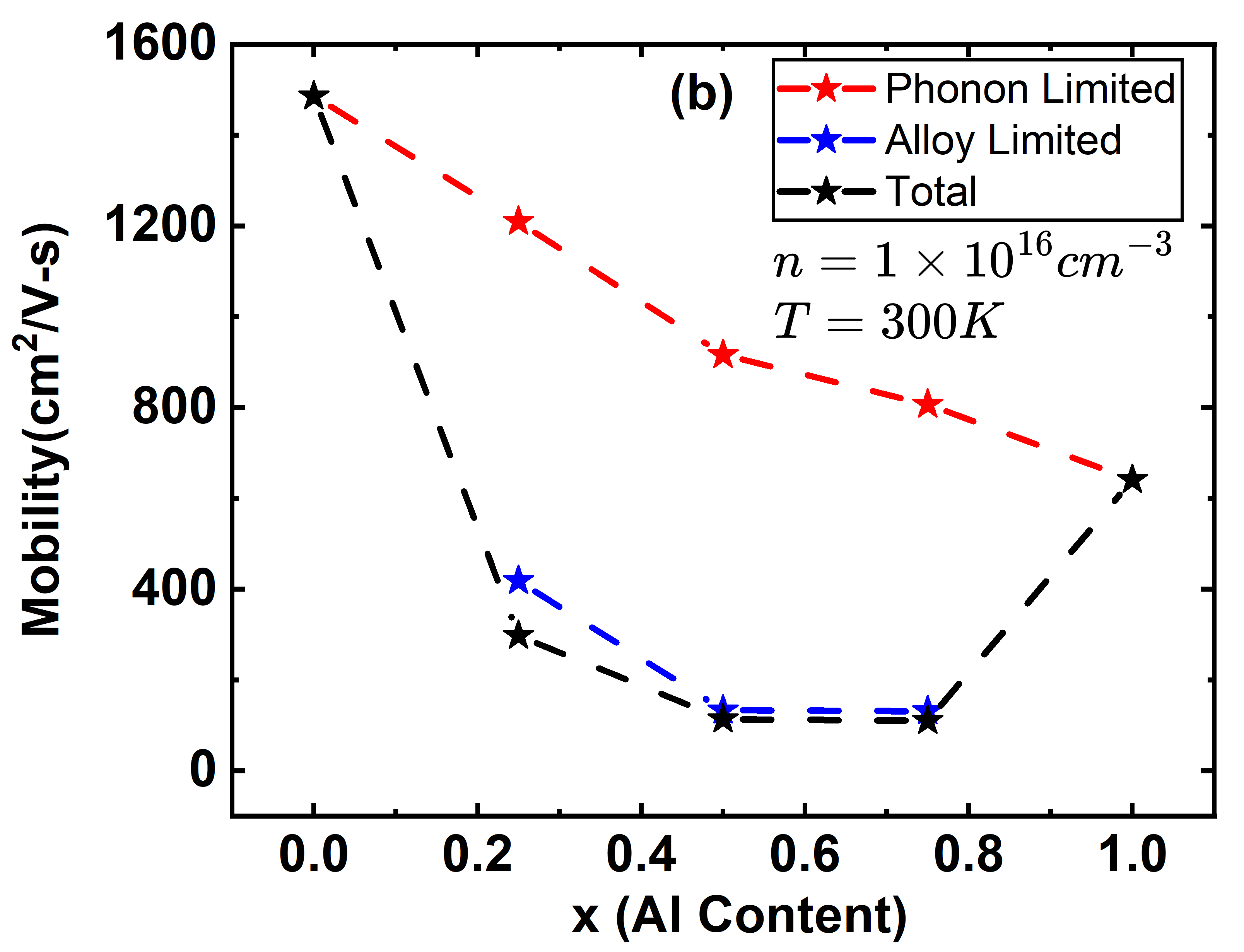}
    \end{minipage}
    \caption{Calculated low field phonon limited, alloy limited, and total mobility for (a)All scattering mechanisms are included (b) Piezoelectric scattering is suppressed for varying Al fraction from GaN to AlN}
    \label{ganalnpedquad}
\end{figure}

The idea discussed in the preceding section can be applied to compute the transport properties in random alloy systems modeled using SC. In this section, the low field transport properties in Al$_x$Ga$_{1-x}$N are investigated for different Al fractions. For the transport calculations, the 24-atom SC is generated from the 10-atom PC as the fundamental repeating unit in the $3\times2\times1$ configuration. To overcome the computational demands associated with extensive simulations, the dimensions of the supercell were reduced from 40 to 24.
\\
In our low-field transport calculations for alloys, the alloy scattering mechanism is included in addition to polar optical phonon scattering, piezoelectric scattering, deformation potential scattering, and ionized impurity scattering. Using the concept discussed in the previous section, the polar optical phonon scattering rate is calculated for an electron concentration of 5.5e16 cm$^{-3}$ at room temperature as shown in Fig.\ref{alganscat}
The alloy disorder scattering is modeled using an analytical expression of Fermi's Golden Rule where the effective scattering potential is used from previously reported theoretical and experimental data \cite{bellotti2007alloy,pant2020high,jena2003magnetotransport,simon2006carrier}.
Then using Rode's iterative method, the phonon scattering limited, alloy scattering limited, and the total low field mobility is calculated at different Al fractions. Fig.\ref{ganalnpedquad} shows the calculated low field mobility for various Al fractions ranging from GaN to AlN at room temperature for an electron concentration of n=1e16 cm$^{-3}$.
In Fig.\ref{ganalnpedquad}a, the phonon limited mobility is calculated taking into account all the scattering mechanisms. However, Fig.\ref{ganalnpedquad}b corresponds to the case where dynamic quadrupole interaction suppresses the piezoelectric scattering in group III nitrides thus making polar optical phonon scattering the dominant scattering mechanism. In this calculation, we have completely suppressed the piezoelectric scattering. This phenomenon is already studied in GaN\cite{jhalani2020piezoelectric} and is applicable in piezoelectric materials like group III-nitrides. It can be seen from Fig.\ref{ganalnpedquad}, that the phonon limited increases significantly when the piezoelectric scattering is suppressed which has already been observed in GaN and recent experimental reports have also demonstrated achieving similar high mobility values\cite{kaneki2024record}. There have been no reports of how the dynamic quadrupoles affect the electron-phonon interaction for AlGaN alloys as well as in AlN. Thus we have completely ignored the effect of piezoelectric scattering to find the maximum limit of phonon limited mobility in AlGaN alloys as well as in AlN. Our calculations show a room temperature mobility of around 700 cm$^2$/V-s for AlN provides an upper limit to the mobility and is overestimated compared to experimental data\cite{bagheri2022high,taniyasu2006increased} which can also include defect scattering.
\\
As reported in earlier studies, the alloy scattering limited mobility follows a bathtub-like nature, where the alloy limited scattering mobility reaches a minimum at Al concentrations $\approx$ 0.6. Our calculations show that the phonon scattering limited mobility decreases monotonically and thus the alloy scattering mechanism is dominant near Al fractions of x$\approx$0.6. However, at other Al fractions closer to GaN and AlN, the polar optical phonon scattering plays an important role along with alloy disorder scattering to limit the low field mobility in AlGaN alloys at room temperature conditions.
\\
 When piezoelectric scattering is suppressed in AlGaN, there is a marked rise in phonon-limited mobility, with alloy scattering emerging as the dominant mechanism across all fractions of Al at room temperature. Since suppressing the piezoelectric scattering gives an upper limit to the phonon limited mobility, in our further discussions we have taken this case to study the dependence of mobility on temperature.
\begin{figure}
\begin{center}
    \includegraphics[width=\linewidth]{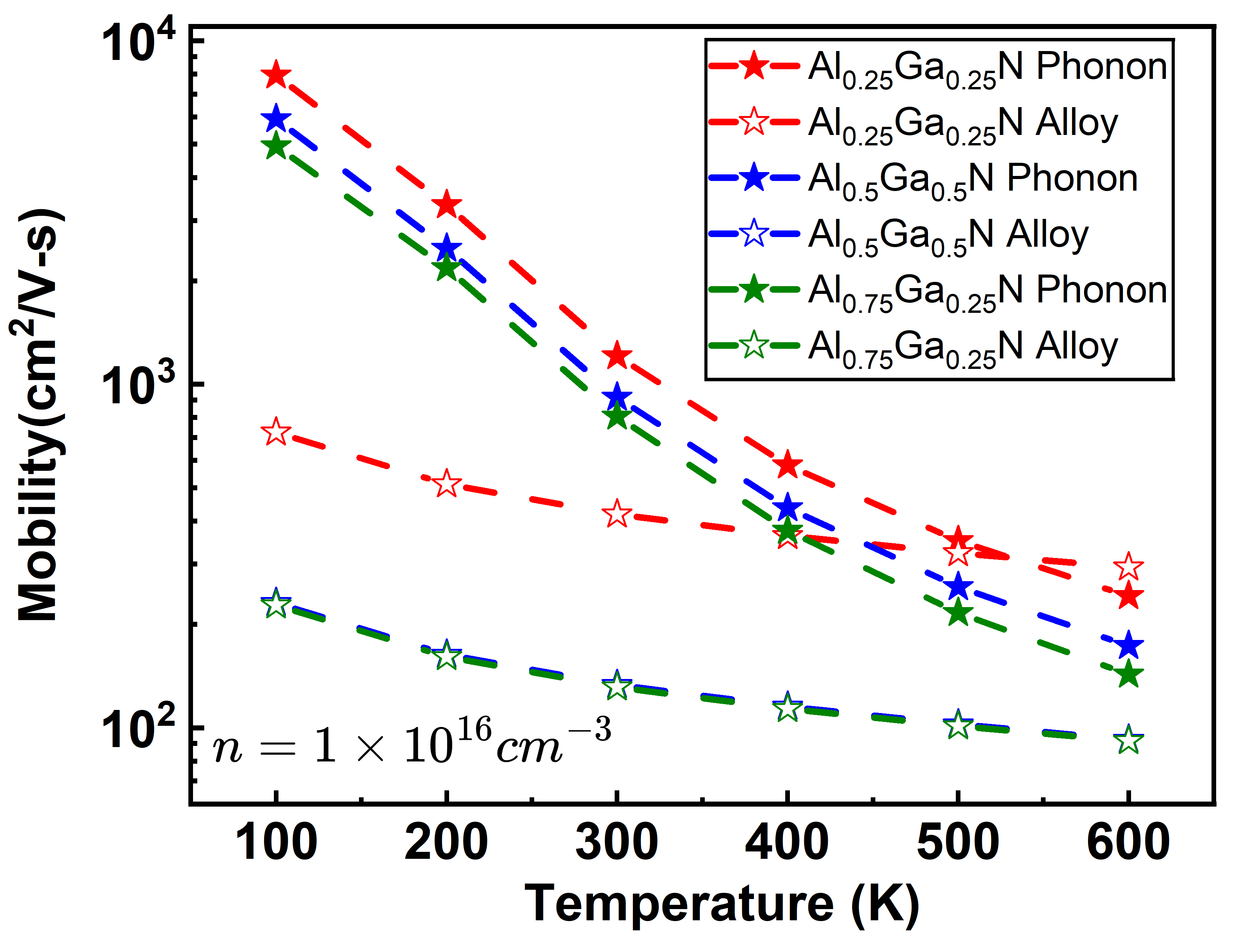}
    \caption{Variation of calculated low field phonon limited and alloy limited mobility with temperature for various Al fractions}
    \label{temptotal}
    \end{center}
\end{figure}
\\
Fig.\ref{temptotal} shows the variation of phonon limited and alloy limited mobility with temperature. At low temperatures phonon scattering is insignificant due to freezing and alloy scattering is the dominant scattering mechanism. However, with increasing temperatures, phonon scattering plays an important role. As observed in Fig.\ref{temptotal}, for 25$\%$ Al fraction, phonon scattering is more dominant than alloy scattering at high temperatures of above 400K. For higher Al fractions, the phonon scattering and alloy scattering limited mobility are comparable at high temperatures of above 400K. Thus phonon scattering mechanisms play an important role in device operations at higher temperatures along with alloy scattering.
\section{Conclusion}
In this work, the effective phonon dispersion (EPD) of AlGaN alloys was investigated using a supercell approach. The individual phonon modes were identified which are in agreement with previously reported experimental data. We have also developed a method to calculate the low-field mobility in supercells using a modified q-grid approach. Our calculations for GaN supercells show good agreement with GaN primitive cell calculations which validates our method. This technique opens up a path to calculate phonon limited mobility from a first principles approach in disordered alloy systems modeled using supercells. The low field transport mechanisms in AlGaN alloys are investigated which shows that along with alloy scattering, electron-phonon scattering mechanisms may also play an important role in limiting the low field mobility at high temperatures.
\section{Acknowledgments}
We acknowledge the support from the ARO (Program Managers: Dr. Joe Qiu and Dr. Tom Oder), AFOSR (Prgram manager: Dr. Ali Sayir) and the Center for Computational Research (CCR) at the University at Buffalo
\bibliography{aipsamp}

\begin{thebibliography}{47}%
\makeatletter
\providecommand \@ifxundefined [1]{%
 \@ifx{#1\undefined}
}%
\providecommand \@ifnum [1]{%
 \ifnum #1\expandafter \@firstoftwo
 \else \expandafter \@secondoftwo
 \fi
}%
\providecommand \@ifx [1]{%
 \ifx #1\expandafter \@firstoftwo
 \else \expandafter \@secondoftwo
 \fi
}%
\providecommand \natexlab [1]{#1}%
\providecommand \enquote  [1]{``#1''}%
\providecommand \bibnamefont  [1]{#1}%
\providecommand \bibfnamefont [1]{#1}%
\providecommand \citenamefont [1]{#1}%
\providecommand \href@noop [0]{\@secondoftwo}%
\providecommand \href [0]{\begingroup \@sanitize@url \@href}%
\providecommand \@href[1]{\@@startlink{#1}\@@href}%
\providecommand \@@href[1]{\endgroup#1\@@endlink}%
\providecommand \@sanitize@url [0]{\catcode `\\12\catcode `\$12\catcode `\&12\catcode `\#12\catcode `\^12\catcode `\_12\catcode `\%12\relax}%
\providecommand \@@startlink[1]{}%
\providecommand \@@endlink[0]{}%
\providecommand \url  [0]{\begingroup\@sanitize@url \@url }%
\providecommand \@url [1]{\endgroup\@href {#1}{\urlprefix }}%
\providecommand \urlprefix  [0]{URL }%
\providecommand \Eprint [0]{\href }%
\providecommand \doibase [0]{http://dx.doi.org/}%
\providecommand \selectlanguage [0]{\@gobble}%
\providecommand \bibinfo  [0]{\@secondoftwo}%
\providecommand \bibfield  [0]{\@secondoftwo}%
\providecommand \translation [1]{[#1]}%
\providecommand \BibitemOpen [0]{}%
\providecommand \bibitemStop [0]{}%
\providecommand \bibitemNoStop [0]{.\EOS\space}%
\providecommand \EOS [0]{\spacefactor3000\relax}%
\providecommand \BibitemShut  [1]{\csname bibitem#1\endcsname}%
\let\auto@bib@innerbib\@empty
\bibitem [{\citenamefont {Casady}\ and\ \citenamefont {Johnson}(1996)}]{casady1996status}%
  \BibitemOpen
  \bibfield  {author} {\bibinfo {author} {\bibfnamefont {J.}~\bibnamefont {Casady}}\ and\ \bibinfo {author} {\bibfnamefont {R.~W.}\ \bibnamefont {Johnson}},\ }\bibfield  {title} {\enquote {\bibinfo {title} {Status of silicon carbide (sic) as a wide-bandgap semiconductor for high-temperature applications: A review},}\ }\href@noop {} {\bibfield  {journal} {\bibinfo  {journal} {Solid-State Electronics}\ }\textbf {\bibinfo {volume} {39}},\ \bibinfo {pages} {1409--1422} (\bibinfo {year} {1996})}\BibitemShut {NoStop}%
\bibitem [{\citenamefont {Nakamura}\ and\ \citenamefont {Fasol}(2013)}]{nakamura2013blue}%
  \BibitemOpen
  \bibfield  {author} {\bibinfo {author} {\bibfnamefont {S.}~\bibnamefont {Nakamura}}\ and\ \bibinfo {author} {\bibfnamefont {G.}~\bibnamefont {Fasol}},\ }\href@noop {} {\emph {\bibinfo {title} {The blue laser diode: GaN based light emitters and lasers}}}\ (\bibinfo  {publisher} {Springer Science \& Business Media},\ \bibinfo {year} {2013})\BibitemShut {NoStop}%
\bibitem [{\citenamefont {Morkoc}\ and\ \citenamefont {Morkoc}(1999)}]{morkocc1999general}%
  \BibitemOpen
  \bibfield  {author} {\bibinfo {author} {\bibfnamefont {H.}~\bibnamefont {Morkoc}}\ and\ \bibinfo {author} {\bibfnamefont {H.}~\bibnamefont {Morkoc}},\ }\bibfield  {title} {\enquote {\bibinfo {title} {General properties of nitrides},}\ }\href@noop {} {\bibfield  {journal} {\bibinfo  {journal} {Nitride semiconductors and devices}\ ,\ \bibinfo {pages} {8--44}} (\bibinfo {year} {1999})}\BibitemShut {NoStop}%
\bibitem [{\citenamefont {Vurgaftman}\ and\ \citenamefont {Meyer}(2003)}]{vurgaftman2003band}%
  \BibitemOpen
  \bibfield  {author} {\bibinfo {author} {\bibfnamefont {I.}~\bibnamefont {Vurgaftman}}\ and\ \bibinfo {author} {\bibfnamefont {J.~n.}\ \bibnamefont {Meyer}},\ }\bibfield  {title} {\enquote {\bibinfo {title} {Band parameters for nitrogen-containing semiconductors},}\ }\href@noop {} {\bibfield  {journal} {\bibinfo  {journal} {Journal of Applied Physics}\ }\textbf {\bibinfo {volume} {94}},\ \bibinfo {pages} {3675--3696} (\bibinfo {year} {2003})}\BibitemShut {NoStop}%
\bibitem [{\citenamefont {Khan}\ \emph {et~al.}(1991)\citenamefont {Khan}, \citenamefont {Olson}, \citenamefont {Van~Hove},\ and\ \citenamefont {Kuznia}}]{khan1991vertical}%
  \BibitemOpen
  \bibfield  {author} {\bibinfo {author} {\bibfnamefont {M.~A.}\ \bibnamefont {Khan}}, \bibinfo {author} {\bibfnamefont {D.}~\bibnamefont {Olson}}, \bibinfo {author} {\bibfnamefont {J.}~\bibnamefont {Van~Hove}}, \ and\ \bibinfo {author} {\bibfnamefont {J.}~\bibnamefont {Kuznia}},\ }\bibfield  {title} {\enquote {\bibinfo {title} {Vertical-cavity, room-temperature stimulated emission from photopumped gan films deposited over sapphire substrates using low-pressure metalorganic chemical vapor deposition},}\ }\href@noop {} {\bibfield  {journal} {\bibinfo  {journal} {Applied Physics Letters}\ }\textbf {\bibinfo {volume} {58}},\ \bibinfo {pages} {1515--1517} (\bibinfo {year} {1991})}\BibitemShut {NoStop}%
\bibitem [{\citenamefont {Higashiwaki}\ \emph {et~al.}(2014)\citenamefont {Higashiwaki}, \citenamefont {Sasaki}, \citenamefont {Kuramata}, \citenamefont {Masui},\ and\ \citenamefont {Yamakoshi}}]{higashiwaki2014development}%
  \BibitemOpen
  \bibfield  {author} {\bibinfo {author} {\bibfnamefont {M.}~\bibnamefont {Higashiwaki}}, \bibinfo {author} {\bibfnamefont {K.}~\bibnamefont {Sasaki}}, \bibinfo {author} {\bibfnamefont {A.}~\bibnamefont {Kuramata}}, \bibinfo {author} {\bibfnamefont {T.}~\bibnamefont {Masui}}, \ and\ \bibinfo {author} {\bibfnamefont {S.}~\bibnamefont {Yamakoshi}},\ }\bibfield  {title} {\enquote {\bibinfo {title} {Development of gallium oxide power devices},}\ }\href@noop {} {\bibfield  {journal} {\bibinfo  {journal} {physica status solidi (a)}\ }\textbf {\bibinfo {volume} {211}},\ \bibinfo {pages} {21--26} (\bibinfo {year} {2014})}\BibitemShut {NoStop}%
\bibitem [{\citenamefont {Wang}\ \emph {et~al.}(2018)\citenamefont {Wang}, \citenamefont {Li}, \citenamefont {Ni},\ and\ \citenamefont {Janotti}}]{wang2018band}%
  \BibitemOpen
  \bibfield  {author} {\bibinfo {author} {\bibfnamefont {T.}~\bibnamefont {Wang}}, \bibinfo {author} {\bibfnamefont {W.}~\bibnamefont {Li}}, \bibinfo {author} {\bibfnamefont {C.}~\bibnamefont {Ni}}, \ and\ \bibinfo {author} {\bibfnamefont {A.}~\bibnamefont {Janotti}},\ }\bibfield  {title} {\enquote {\bibinfo {title} {Band gap and band offset of ${Ga}_2{O}_3$ and $({Al}_x{Ga}_{1-x})_2{O}_3$ alloys},}\ }\href@noop {} {\bibfield  {journal} {\bibinfo  {journal} {Physical Review Applied}\ }\textbf {\bibinfo {volume} {10}},\ \bibinfo {pages} {011003} (\bibinfo {year} {2018})}\BibitemShut {NoStop}%
\bibitem [{\citenamefont {Khan}, \citenamefont {Balakrishnan},\ and\ \citenamefont {Katona}(2008)}]{khan2008ultraviolet}%
  \BibitemOpen
  \bibfield  {author} {\bibinfo {author} {\bibfnamefont {A.}~\bibnamefont {Khan}}, \bibinfo {author} {\bibfnamefont {K.}~\bibnamefont {Balakrishnan}}, \ and\ \bibinfo {author} {\bibfnamefont {T.}~\bibnamefont {Katona}},\ }\bibfield  {title} {\enquote {\bibinfo {title} {Ultraviolet light-emitting diodes based on group three nitrides},}\ }\href@noop {} {\bibfield  {journal} {\bibinfo  {journal} {Nature photonics}\ }\textbf {\bibinfo {volume} {2}},\ \bibinfo {pages} {77--84} (\bibinfo {year} {2008})}\BibitemShut {NoStop}%
\bibitem [{\citenamefont {Pimputkar}\ \emph {et~al.}(2009)\citenamefont {Pimputkar}, \citenamefont {Speck}, \citenamefont {DenBaars},\ and\ \citenamefont {Nakamura}}]{pimputkar2009prospects}%
  \BibitemOpen
  \bibfield  {author} {\bibinfo {author} {\bibfnamefont {S.}~\bibnamefont {Pimputkar}}, \bibinfo {author} {\bibfnamefont {J.~S.}\ \bibnamefont {Speck}}, \bibinfo {author} {\bibfnamefont {S.~P.}\ \bibnamefont {DenBaars}}, \ and\ \bibinfo {author} {\bibfnamefont {S.}~\bibnamefont {Nakamura}},\ }\bibfield  {title} {\enquote {\bibinfo {title} {Prospects for led lighting},}\ }\href@noop {} {\bibfield  {journal} {\bibinfo  {journal} {Nature photonics}\ }\textbf {\bibinfo {volume} {3}},\ \bibinfo {pages} {180--182} (\bibinfo {year} {2009})}\BibitemShut {NoStop}%
\bibitem [{\citenamefont {Mishra}, \citenamefont {Parikh},\ and\ \citenamefont {Wu}(2002)}]{mishra2002algan}%
  \BibitemOpen
  \bibfield  {author} {\bibinfo {author} {\bibfnamefont {U.~K.}\ \bibnamefont {Mishra}}, \bibinfo {author} {\bibfnamefont {P.}~\bibnamefont {Parikh}}, \ and\ \bibinfo {author} {\bibfnamefont {Y.-F.}\ \bibnamefont {Wu}},\ }\bibfield  {title} {\enquote {\bibinfo {title} {{Al}{Ga}{N}/{Ga}{N} hemts-an overview of device operation and applications},}\ }\href@noop {} {\bibfield  {journal} {\bibinfo  {journal} {Proceedings of the IEEE}\ }\textbf {\bibinfo {volume} {90}},\ \bibinfo {pages} {1022--1031} (\bibinfo {year} {2002})}\BibitemShut {NoStop}%
\bibitem [{\citenamefont {Mishra}\ \emph {et~al.}(2008)\citenamefont {Mishra}, \citenamefont {Shen}, \citenamefont {Kazior},\ and\ \citenamefont {Wu}}]{mishra2008gan}%
  \BibitemOpen
  \bibfield  {author} {\bibinfo {author} {\bibfnamefont {U.~K.}\ \bibnamefont {Mishra}}, \bibinfo {author} {\bibfnamefont {L.}~\bibnamefont {Shen}}, \bibinfo {author} {\bibfnamefont {T.~E.}\ \bibnamefont {Kazior}}, \ and\ \bibinfo {author} {\bibfnamefont {Y.-F.}\ \bibnamefont {Wu}},\ }\bibfield  {title} {\enquote {\bibinfo {title} {Gan-based rf power devices and amplifiers},}\ }\href@noop {} {\bibfield  {journal} {\bibinfo  {journal} {Proceedings of the IEEE}\ }\textbf {\bibinfo {volume} {96}},\ \bibinfo {pages} {287--305} (\bibinfo {year} {2008})}\BibitemShut {NoStop}%
\bibitem [{\citenamefont {Strite}\ and\ \citenamefont {Morkoc}(1992)}]{strite1992gan}%
  \BibitemOpen
  \bibfield  {author} {\bibinfo {author} {\bibfnamefont {a.}~\bibnamefont {Strite}}\ and\ \bibinfo {author} {\bibfnamefont {H.}~\bibnamefont {Morkoc}},\ }\bibfield  {title} {\enquote {\bibinfo {title} {Gan, aln, and inn: a review},}\ }\href@noop {} {\bibfield  {journal} {\bibinfo  {journal} {Journal of Vacuum Science \& Technology B: Microelectronics and Nanometer Structures Processing, Measurement, and Phenomena}\ }\textbf {\bibinfo {volume} {10}},\ \bibinfo {pages} {1237--1266} (\bibinfo {year} {1992})}\BibitemShut {NoStop}%
\bibitem [{\citenamefont {Kyrtsos}, \citenamefont {Matsubara},\ and\ \citenamefont {Bellotti}(2019)}]{kyrtsos2019first}%
  \BibitemOpen
  \bibfield  {author} {\bibinfo {author} {\bibfnamefont {A.}~\bibnamefont {Kyrtsos}}, \bibinfo {author} {\bibfnamefont {M.}~\bibnamefont {Matsubara}}, \ and\ \bibinfo {author} {\bibfnamefont {E.}~\bibnamefont {Bellotti}},\ }\bibfield  {title} {\enquote {\bibinfo {title} {First-principles study of the impact of the atomic configuration on the electronic properties of ${Al}_x{Ga}_{1- x}{N}$ alloys},}\ }\href@noop {} {\bibfield  {journal} {\bibinfo  {journal} {Physical Review B}\ }\textbf {\bibinfo {volume} {99}},\ \bibinfo {pages} {035201} (\bibinfo {year} {2019})}\BibitemShut {NoStop}%
\bibitem [{\citenamefont {Coltrin}\ and\ \citenamefont {Kaplar}(2017)}]{coltrin2017transport}%
  \BibitemOpen
  \bibfield  {author} {\bibinfo {author} {\bibfnamefont {M.~E.}\ \bibnamefont {Coltrin}}\ and\ \bibinfo {author} {\bibfnamefont {R.~J.}\ \bibnamefont {Kaplar}},\ }\bibfield  {title} {\enquote {\bibinfo {title} {Transport and breakdown analysis for improved figure-of-merit for {Al}{Ga}{N} power devices},}\ }\href@noop {} {\bibfield  {journal} {\bibinfo  {journal} {Journal of Applied Physics}\ }\textbf {\bibinfo {volume} {121}} (\bibinfo {year} {2017})}\BibitemShut {NoStop}%
\bibitem [{\citenamefont {Pant}, \citenamefont {Deng},\ and\ \citenamefont {Kioupakis}(2020)}]{pant2020high}%
  \BibitemOpen
  \bibfield  {author} {\bibinfo {author} {\bibfnamefont {N.}~\bibnamefont {Pant}}, \bibinfo {author} {\bibfnamefont {Z.}~\bibnamefont {Deng}}, \ and\ \bibinfo {author} {\bibfnamefont {E.}~\bibnamefont {Kioupakis}},\ }\bibfield  {title} {\enquote {\bibinfo {title} {High electron mobility of ${Al}_x{Ga}_{1- x}{N}$ evaluated by unfolding the dft band structure},}\ }\href@noop {} {\bibfield  {journal} {\bibinfo  {journal} {Applied Physics Letters}\ }\textbf {\bibinfo {volume} {117}} (\bibinfo {year} {2020})}\BibitemShut {NoStop}%
\bibitem [{\citenamefont {Bellotti}, \citenamefont {Bertazzi},\ and\ \citenamefont {Goano}(2007)}]{bellotti2007alloy}%
  \BibitemOpen
  \bibfield  {author} {\bibinfo {author} {\bibfnamefont {E.}~\bibnamefont {Bellotti}}, \bibinfo {author} {\bibfnamefont {F.}~\bibnamefont {Bertazzi}}, \ and\ \bibinfo {author} {\bibfnamefont {M.}~\bibnamefont {Goano}},\ }\bibfield  {title} {\enquote {\bibinfo {title} {Alloy scattering in {Al}{Ga}{N} and {In}{Ga}{N}: A numerical study},}\ }\href@noop {} {\bibfield  {journal} {\bibinfo  {journal} {Journal of Applied Physics}\ }\textbf {\bibinfo {volume} {101}} (\bibinfo {year} {2007})}\BibitemShut {NoStop}%
\bibitem [{\citenamefont {Farahmand}\ \emph {et~al.}(2001)\citenamefont {Farahmand}, \citenamefont {Garetto}, \citenamefont {Bellotti}, \citenamefont {Brennan}, \citenamefont {Goano}, \citenamefont {Ghillino}, \citenamefont {Ghione}, \citenamefont {Albrecht},\ and\ \citenamefont {Ruden}}]{farahmand2001monte}%
  \BibitemOpen
  \bibfield  {author} {\bibinfo {author} {\bibfnamefont {M.}~\bibnamefont {Farahmand}}, \bibinfo {author} {\bibfnamefont {C.}~\bibnamefont {Garetto}}, \bibinfo {author} {\bibfnamefont {E.}~\bibnamefont {Bellotti}}, \bibinfo {author} {\bibfnamefont {K.~F.}\ \bibnamefont {Brennan}}, \bibinfo {author} {\bibfnamefont {M.}~\bibnamefont {Goano}}, \bibinfo {author} {\bibfnamefont {E.}~\bibnamefont {Ghillino}}, \bibinfo {author} {\bibfnamefont {G.}~\bibnamefont {Ghione}}, \bibinfo {author} {\bibfnamefont {J.~D.}\ \bibnamefont {Albrecht}}, \ and\ \bibinfo {author} {\bibfnamefont {P.~P.}\ \bibnamefont {Ruden}},\ }\bibfield  {title} {\enquote {\bibinfo {title} {Monte carlo simulation of electron transport in the iii-nitride wurtzite phase materials system: binaries and ternaries},}\ }\href@noop {} {\bibfield  {journal} {\bibinfo  {journal} {IEEE Transactions on electron devices}\ }\textbf {\bibinfo {volume} {48}},\ \bibinfo {pages} {535--542} (\bibinfo {year} {2001})}\BibitemShut {NoStop}%
\bibitem [{\citenamefont {Zunger}\ \emph {et~al.}(1990)\citenamefont {Zunger}, \citenamefont {Wei}, \citenamefont {Ferreira},\ and\ \citenamefont {Bernard}}]{zunger1990special}%
  \BibitemOpen
  \bibfield  {author} {\bibinfo {author} {\bibfnamefont {A.}~\bibnamefont {Zunger}}, \bibinfo {author} {\bibfnamefont {S.-H.}\ \bibnamefont {Wei}}, \bibinfo {author} {\bibfnamefont {L.}~\bibnamefont {Ferreira}}, \ and\ \bibinfo {author} {\bibfnamefont {J.~E.}\ \bibnamefont {Bernard}},\ }\bibfield  {title} {\enquote {\bibinfo {title} {Special quasirandom structures},}\ }\href@noop {} {\bibfield  {journal} {\bibinfo  {journal} {Physical review letters}\ }\textbf {\bibinfo {volume} {65}},\ \bibinfo {pages} {353} (\bibinfo {year} {1990})}\BibitemShut {NoStop}%
\bibitem [{\citenamefont {Wei}\ \emph {et~al.}(1990)\citenamefont {Wei}, \citenamefont {Ferreira}, \citenamefont {Bernard},\ and\ \citenamefont {Zunger}}]{wei1990electronic}%
  \BibitemOpen
  \bibfield  {author} {\bibinfo {author} {\bibfnamefont {S.-H.}\ \bibnamefont {Wei}}, \bibinfo {author} {\bibfnamefont {L.}~\bibnamefont {Ferreira}}, \bibinfo {author} {\bibfnamefont {J.~E.}\ \bibnamefont {Bernard}}, \ and\ \bibinfo {author} {\bibfnamefont {A.}~\bibnamefont {Zunger}},\ }\bibfield  {title} {\enquote {\bibinfo {title} {Electronic properties of random alloys: Special quasirandom structures},}\ }\href@noop {} {\bibfield  {journal} {\bibinfo  {journal} {Physical Review B}\ }\textbf {\bibinfo {volume} {42}},\ \bibinfo {pages} {9622} (\bibinfo {year} {1990})}\BibitemShut {NoStop}%
\bibitem [{\citenamefont {Wang}\ \emph {et~al.}(1998)\citenamefont {Wang}, \citenamefont {Bellaiche}, \citenamefont {Wei},\ and\ \citenamefont {Zunger}}]{wang1998majority}%
  \BibitemOpen
  \bibfield  {author} {\bibinfo {author} {\bibfnamefont {L.-W.}\ \bibnamefont {Wang}}, \bibinfo {author} {\bibfnamefont {L.}~\bibnamefont {Bellaiche}}, \bibinfo {author} {\bibfnamefont {S.-H.}\ \bibnamefont {Wei}}, \ and\ \bibinfo {author} {\bibfnamefont {A.}~\bibnamefont {Zunger}},\ }\bibfield  {title} {\enquote {\bibinfo {title} {“majority representation” of alloy electronic states},}\ }\href@noop {} {\bibfield  {journal} {\bibinfo  {journal} {Physical review letters}\ }\textbf {\bibinfo {volume} {80}},\ \bibinfo {pages} {4725} (\bibinfo {year} {1998})}\BibitemShut {NoStop}%
\bibitem [{\citenamefont {Van~de Walle}\ \emph {et~al.}(2013)\citenamefont {Van~de Walle}, \citenamefont {Tiwary}, \citenamefont {De~Jong}, \citenamefont {Olmsted}, \citenamefont {Asta}, \citenamefont {Dick}, \citenamefont {Shin}, \citenamefont {Wang}, \citenamefont {Chen},\ and\ \citenamefont {Liu}}]{van2013efficient}%
  \BibitemOpen
  \bibfield  {author} {\bibinfo {author} {\bibfnamefont {A.}~\bibnamefont {Van~de Walle}}, \bibinfo {author} {\bibfnamefont {P.}~\bibnamefont {Tiwary}}, \bibinfo {author} {\bibfnamefont {M.}~\bibnamefont {De~Jong}}, \bibinfo {author} {\bibfnamefont {D.}~\bibnamefont {Olmsted}}, \bibinfo {author} {\bibfnamefont {M.}~\bibnamefont {Asta}}, \bibinfo {author} {\bibfnamefont {A.}~\bibnamefont {Dick}}, \bibinfo {author} {\bibfnamefont {D.}~\bibnamefont {Shin}}, \bibinfo {author} {\bibfnamefont {Y.}~\bibnamefont {Wang}}, \bibinfo {author} {\bibfnamefont {L.-Q.}\ \bibnamefont {Chen}}, \ and\ \bibinfo {author} {\bibfnamefont {Z.-K.}\ \bibnamefont {Liu}},\ }\bibfield  {title} {\enquote {\bibinfo {title} {Efficient stochastic generation of special quasirandom structures},}\ }\href@noop {} {\bibfield  {journal} {\bibinfo  {journal} {Calphad}\ }\textbf {\bibinfo {volume} {42}},\ \bibinfo {pages} {13--18} (\bibinfo {year} {2013})}\BibitemShut {NoStop}%
\bibitem [{\citenamefont {Van De~Walle}(2009)}]{van2009multicomponent}%
  \BibitemOpen
  \bibfield  {author} {\bibinfo {author} {\bibfnamefont {A.}~\bibnamefont {Van De~Walle}},\ }\bibfield  {title} {\enquote {\bibinfo {title} {Multicomponent multisublattice alloys, nonconfigurational entropy and other additions to the alloy theoretic automated toolkit},}\ }\href@noop {} {\bibfield  {journal} {\bibinfo  {journal} {Calphad}\ }\textbf {\bibinfo {volume} {33}},\ \bibinfo {pages} {266--278} (\bibinfo {year} {2009})}\BibitemShut {NoStop}%
\bibitem [{\citenamefont {Giannozzi}\ \emph {et~al.}(2009)\citenamefont {Giannozzi}, \citenamefont {Baroni}, \citenamefont {Bonini}, \citenamefont {Calandra}, \citenamefont {Car}, \citenamefont {Cavazzoni}, \citenamefont {Ceresoli}, \citenamefont {Chiarotti}, \citenamefont {Cococcioni}, \citenamefont {Dabo} \emph {et~al.}}]{giannozzi2009quantum}%
  \BibitemOpen
  \bibfield  {author} {\bibinfo {author} {\bibfnamefont {P.}~\bibnamefont {Giannozzi}}, \bibinfo {author} {\bibfnamefont {S.}~\bibnamefont {Baroni}}, \bibinfo {author} {\bibfnamefont {N.}~\bibnamefont {Bonini}}, \bibinfo {author} {\bibfnamefont {M.}~\bibnamefont {Calandra}}, \bibinfo {author} {\bibfnamefont {R.}~\bibnamefont {Car}}, \bibinfo {author} {\bibfnamefont {C.}~\bibnamefont {Cavazzoni}}, \bibinfo {author} {\bibfnamefont {D.}~\bibnamefont {Ceresoli}}, \bibinfo {author} {\bibfnamefont {G.~L.}\ \bibnamefont {Chiarotti}}, \bibinfo {author} {\bibfnamefont {M.}~\bibnamefont {Cococcioni}}, \bibinfo {author} {\bibfnamefont {I.}~\bibnamefont {Dabo}},  \emph {et~al.},\ }\bibfield  {title} {\enquote {\bibinfo {title} {Quantum espresso: a modular and open-source software project for quantum simulations of materials},}\ }\href@noop {} {\bibfield  {journal} {\bibinfo  {journal} {Journal of physics: Condensed matter}\ }\textbf {\bibinfo {volume} {21}},\ \bibinfo {pages} {395502} (\bibinfo {year}
  {2009})}\BibitemShut {NoStop}%
\bibitem [{\citenamefont {Giannozzi}\ \emph {et~al.}(2017)\citenamefont {Giannozzi}, \citenamefont {Andreussi}, \citenamefont {Brumme}, \citenamefont {Bunau}, \citenamefont {Nardelli}, \citenamefont {Calandra}, \citenamefont {Car}, \citenamefont {Cavazzoni}, \citenamefont {Ceresoli}, \citenamefont {Cococcioni} \emph {et~al.}}]{giannozzi2017advanced}%
  \BibitemOpen
  \bibfield  {author} {\bibinfo {author} {\bibfnamefont {P.}~\bibnamefont {Giannozzi}}, \bibinfo {author} {\bibfnamefont {O.}~\bibnamefont {Andreussi}}, \bibinfo {author} {\bibfnamefont {T.}~\bibnamefont {Brumme}}, \bibinfo {author} {\bibfnamefont {O.}~\bibnamefont {Bunau}}, \bibinfo {author} {\bibfnamefont {M.~B.}\ \bibnamefont {Nardelli}}, \bibinfo {author} {\bibfnamefont {M.}~\bibnamefont {Calandra}}, \bibinfo {author} {\bibfnamefont {R.}~\bibnamefont {Car}}, \bibinfo {author} {\bibfnamefont {C.}~\bibnamefont {Cavazzoni}}, \bibinfo {author} {\bibfnamefont {D.}~\bibnamefont {Ceresoli}}, \bibinfo {author} {\bibfnamefont {M.}~\bibnamefont {Cococcioni}},  \emph {et~al.},\ }\bibfield  {title} {\enquote {\bibinfo {title} {Advanced capabilities for materials modelling with quantum espresso},}\ }\href@noop {} {\bibfield  {journal} {\bibinfo  {journal} {Journal of physics: Condensed matter}\ }\textbf {\bibinfo {volume} {29}},\ \bibinfo {pages} {465901} (\bibinfo {year} {2017})}\BibitemShut {NoStop}%
\bibitem [{\citenamefont {Gonze}\ and\ \citenamefont {Lee}(1997)}]{gonze1997dynamical}%
  \BibitemOpen
  \bibfield  {author} {\bibinfo {author} {\bibfnamefont {X.}~\bibnamefont {Gonze}}\ and\ \bibinfo {author} {\bibfnamefont {C.}~\bibnamefont {Lee}},\ }\bibfield  {title} {\enquote {\bibinfo {title} {Dynamical matrices, born effective charges, dielectric permittivity tensors, and interatomic force constants from density-functional perturbation theory},}\ }\href@noop {} {\bibfield  {journal} {\bibinfo  {journal} {Physical Review B}\ }\textbf {\bibinfo {volume} {55}},\ \bibinfo {pages} {10355} (\bibinfo {year} {1997})}\BibitemShut {NoStop}%
\bibitem [{\citenamefont {Gonze}(1995)}]{gonze1995adiabatic}%
  \BibitemOpen
  \bibfield  {author} {\bibinfo {author} {\bibfnamefont {X.}~\bibnamefont {Gonze}},\ }\bibfield  {title} {\enquote {\bibinfo {title} {Adiabatic density-functional perturbation theory},}\ }\href@noop {} {\bibfield  {journal} {\bibinfo  {journal} {Physical Review A}\ }\textbf {\bibinfo {volume} {52}},\ \bibinfo {pages} {1096} (\bibinfo {year} {1995})}\BibitemShut {NoStop}%
\bibitem [{\citenamefont {Verdi}\ and\ \citenamefont {Giustino}(2015)}]{verdi2015frohlich}%
  \BibitemOpen
  \bibfield  {author} {\bibinfo {author} {\bibfnamefont {C.}~\bibnamefont {Verdi}}\ and\ \bibinfo {author} {\bibfnamefont {F.}~\bibnamefont {Giustino}},\ }\bibfield  {title} {\enquote {\bibinfo {title} {Frohlich electron-phonon vertex from first principles},}\ }\href@noop {} {\bibfield  {journal} {\bibinfo  {journal} {Physical review letters}\ }\textbf {\bibinfo {volume} {115}},\ \bibinfo {pages} {176401} (\bibinfo {year} {2015})}\BibitemShut {NoStop}%
\bibitem [{\citenamefont {Ponc{\'e}}\ \emph {et~al.}(2016)\citenamefont {Ponc{\'e}}, \citenamefont {Margine}, \citenamefont {Verdi},\ and\ \citenamefont {Giustino}}]{ponce2016epw}%
  \BibitemOpen
  \bibfield  {author} {\bibinfo {author} {\bibfnamefont {S.}~\bibnamefont {Ponc{\'e}}}, \bibinfo {author} {\bibfnamefont {E.~R.}\ \bibnamefont {Margine}}, \bibinfo {author} {\bibfnamefont {C.}~\bibnamefont {Verdi}}, \ and\ \bibinfo {author} {\bibfnamefont {F.}~\bibnamefont {Giustino}},\ }\bibfield  {title} {\enquote {\bibinfo {title} {Epw: Electron--phonon coupling, transport and superconducting properties using maximally localized wannier functions},}\ }\href@noop {} {\bibfield  {journal} {\bibinfo  {journal} {Computer Physics Communications}\ }\textbf {\bibinfo {volume} {209}},\ \bibinfo {pages} {116--133} (\bibinfo {year} {2016})}\BibitemShut {NoStop}%
\bibitem [{\citenamefont {Chattopadhyay}\ and\ \citenamefont {Queisser}(1981)}]{chattopadhyay1981electron}%
  \BibitemOpen
  \bibfield  {author} {\bibinfo {author} {\bibfnamefont {D.}~\bibnamefont {Chattopadhyay}}\ and\ \bibinfo {author} {\bibfnamefont {H.}~\bibnamefont {Queisser}},\ }\bibfield  {title} {\enquote {\bibinfo {title} {Electron scattering by ionized impurities in semiconductors},}\ }\href@noop {} {\bibfield  {journal} {\bibinfo  {journal} {Reviews of Modern Physics}\ }\textbf {\bibinfo {volume} {53}},\ \bibinfo {pages} {745} (\bibinfo {year} {1981})}\BibitemShut {NoStop}%
\bibitem [{\citenamefont {Roccaforte}, \citenamefont {Giannazzo},\ and\ \citenamefont {Greco}(2022)}]{roccaforte2022ion}%
  \BibitemOpen
  \bibfield  {author} {\bibinfo {author} {\bibfnamefont {F.}~\bibnamefont {Roccaforte}}, \bibinfo {author} {\bibfnamefont {F.}~\bibnamefont {Giannazzo}}, \ and\ \bibinfo {author} {\bibfnamefont {G.}~\bibnamefont {Greco}},\ }\bibfield  {title} {\enquote {\bibinfo {title} {Ion implantation doping in silicon carbide and gallium nitride electronic devices},}\ }in\ \href@noop {} {\emph {\bibinfo {booktitle} {Micro}}},\ Vol.~\bibinfo {volume} {2}\ (\bibinfo {organization} {MDPI},\ \bibinfo {year} {2022})\ pp.\ \bibinfo {pages} {23--53}\BibitemShut {NoStop}%
\bibitem [{\citenamefont {Kanechika}\ and\ \citenamefont {Kachi}(2006)}]{kanechika2006n}%
  \BibitemOpen
  \bibfield  {author} {\bibinfo {author} {\bibfnamefont {M.}~\bibnamefont {Kanechika}}\ and\ \bibinfo {author} {\bibfnamefont {T.}~\bibnamefont {Kachi}},\ }\bibfield  {title} {\enquote {\bibinfo {title} {n-type aln layer by si ion implantation},}\ }\href@noop {} {\bibfield  {journal} {\bibinfo  {journal} {Applied physics letters}\ }\textbf {\bibinfo {volume} {88}} (\bibinfo {year} {2006})}\BibitemShut {NoStop}%
\bibitem [{\citenamefont {Mavroidis}\ \emph {et~al.}(2003)\citenamefont {Mavroidis}, \citenamefont {Harris}, \citenamefont {Kappers}, \citenamefont {Humphreys},\ and\ \citenamefont {Bougrioua}}]{mavroidis2003detailed}%
  \BibitemOpen
  \bibfield  {author} {\bibinfo {author} {\bibfnamefont {C.}~\bibnamefont {Mavroidis}}, \bibinfo {author} {\bibfnamefont {J.}~\bibnamefont {Harris}}, \bibinfo {author} {\bibfnamefont {M.}~\bibnamefont {Kappers}}, \bibinfo {author} {\bibfnamefont {C.}~\bibnamefont {Humphreys}}, \ and\ \bibinfo {author} {\bibfnamefont {Z.}~\bibnamefont {Bougrioua}},\ }\bibfield  {title} {\enquote {\bibinfo {title} {Detailed interpretation of electron transport in n-gan},}\ }\href@noop {} {\bibfield  {journal} {\bibinfo  {journal} {Journal of applied physics}\ }\textbf {\bibinfo {volume} {93}},\ \bibinfo {pages} {9095--9103} (\bibinfo {year} {2003})}\BibitemShut {NoStop}%
\bibitem [{\citenamefont {Huang}\ \emph {et~al.}(2001)\citenamefont {Huang}, \citenamefont {Yun}, \citenamefont {Reshchikov}, \citenamefont {Wang}, \citenamefont {Morko{\c{c}}}, \citenamefont {Rode}, \citenamefont {Farina}, \citenamefont {Kurdak}, \citenamefont {Tsen}, \citenamefont {Park} \emph {et~al.}}]{huang2001hall}%
  \BibitemOpen
  \bibfield  {author} {\bibinfo {author} {\bibfnamefont {D.}~\bibnamefont {Huang}}, \bibinfo {author} {\bibfnamefont {F.}~\bibnamefont {Yun}}, \bibinfo {author} {\bibfnamefont {M.}~\bibnamefont {Reshchikov}}, \bibinfo {author} {\bibfnamefont {D.}~\bibnamefont {Wang}}, \bibinfo {author} {\bibfnamefont {H.}~\bibnamefont {Morko{\c{c}}}}, \bibinfo {author} {\bibfnamefont {D.}~\bibnamefont {Rode}}, \bibinfo {author} {\bibfnamefont {L.}~\bibnamefont {Farina}}, \bibinfo {author} {\bibfnamefont {{\c{C}}.}~\bibnamefont {Kurdak}}, \bibinfo {author} {\bibfnamefont {K.-T.}\ \bibnamefont {Tsen}}, \bibinfo {author} {\bibfnamefont {S.}~\bibnamefont {Park}},  \emph {et~al.},\ }\bibfield  {title} {\enquote {\bibinfo {title} {Hall mobility and carrier concentration in free-standing high quality gan templates grown by hydride vapor phase epitaxy},}\ }\href@noop {} {\bibfield  {journal} {\bibinfo  {journal} {Solid-State Electronics}\ }\textbf {\bibinfo {volume} {45}},\ \bibinfo {pages} {711--715} (\bibinfo {year}
  {2001})}\BibitemShut {NoStop}%
\bibitem [{\citenamefont {Xu}\ and\ \citenamefont {Ching}(1993)}]{xu1993electronic}%
  \BibitemOpen
  \bibfield  {author} {\bibinfo {author} {\bibfnamefont {Y.-N.}\ \bibnamefont {Xu}}\ and\ \bibinfo {author} {\bibfnamefont {W.}~\bibnamefont {Ching}},\ }\bibfield  {title} {\enquote {\bibinfo {title} {Electronic, optical, and structural properties of some wurtzite crystals},}\ }\href@noop {} {\bibfield  {journal} {\bibinfo  {journal} {Physical Review B}\ }\textbf {\bibinfo {volume} {48}},\ \bibinfo {pages} {4335} (\bibinfo {year} {1993})}\BibitemShut {NoStop}%
\bibitem [{\citenamefont {Dreyer}, \citenamefont {Janotti},\ and\ \citenamefont {Van~de Walle}(2013)}]{dreyer2013effects}%
  \BibitemOpen
  \bibfield  {author} {\bibinfo {author} {\bibfnamefont {C.}~\bibnamefont {Dreyer}}, \bibinfo {author} {\bibfnamefont {A.}~\bibnamefont {Janotti}}, \ and\ \bibinfo {author} {\bibfnamefont {C.}~\bibnamefont {Van~de Walle}},\ }\bibfield  {title} {\enquote {\bibinfo {title} {Effects of strain on the electron effective mass in gan and aln},}\ }\href@noop {} {\bibfield  {journal} {\bibinfo  {journal} {Applied Physics Letters}\ }\textbf {\bibinfo {volume} {102}} (\bibinfo {year} {2013})}\BibitemShut {NoStop}%
\bibitem [{\citenamefont {Boykin}\ \emph {et~al.}(2014)\citenamefont {Boykin}, \citenamefont {Ajoy}, \citenamefont {Ilatikhameneh}, \citenamefont {Povolotskyi},\ and\ \citenamefont {Klimeck}}]{boykin2014brillouin}%
  \BibitemOpen
  \bibfield  {author} {\bibinfo {author} {\bibfnamefont {T.~B.}\ \bibnamefont {Boykin}}, \bibinfo {author} {\bibfnamefont {A.}~\bibnamefont {Ajoy}}, \bibinfo {author} {\bibfnamefont {H.}~\bibnamefont {Ilatikhameneh}}, \bibinfo {author} {\bibfnamefont {M.}~\bibnamefont {Povolotskyi}}, \ and\ \bibinfo {author} {\bibfnamefont {G.}~\bibnamefont {Klimeck}},\ }\bibfield  {title} {\enquote {\bibinfo {title} {Brillouin zone unfolding method for effective phonon spectra},}\ }\href@noop {} {\bibfield  {journal} {\bibinfo  {journal} {Physical Review B}\ }\textbf {\bibinfo {volume} {90}},\ \bibinfo {pages} {205214} (\bibinfo {year} {2014})}\BibitemShut {NoStop}%
\bibitem [{\citenamefont {Holtz}\ \emph {et~al.}(2001)\citenamefont {Holtz}, \citenamefont {Prokofyeva}, \citenamefont {Seon}, \citenamefont {Copeland}, \citenamefont {Vanbuskirk}, \citenamefont {Williams}, \citenamefont {Nikishin}, \citenamefont {Tretyakov},\ and\ \citenamefont {Temkin}}]{holtz2001composition}%
  \BibitemOpen
  \bibfield  {author} {\bibinfo {author} {\bibfnamefont {M.}~\bibnamefont {Holtz}}, \bibinfo {author} {\bibfnamefont {T.}~\bibnamefont {Prokofyeva}}, \bibinfo {author} {\bibfnamefont {M.}~\bibnamefont {Seon}}, \bibinfo {author} {\bibfnamefont {K.}~\bibnamefont {Copeland}}, \bibinfo {author} {\bibfnamefont {J.}~\bibnamefont {Vanbuskirk}}, \bibinfo {author} {\bibfnamefont {S.}~\bibnamefont {Williams}}, \bibinfo {author} {\bibfnamefont {S.}~\bibnamefont {Nikishin}}, \bibinfo {author} {\bibfnamefont {V.}~\bibnamefont {Tretyakov}}, \ and\ \bibinfo {author} {\bibfnamefont {H.}~\bibnamefont {Temkin}},\ }\bibfield  {title} {\enquote {\bibinfo {title} {Composition dependence of the optical phonon energies in hexagonal ${Al}_x{Ga}_{1- x}{N}$},}\ }\href@noop {} {\bibfield  {journal} {\bibinfo  {journal} {Journal of applied Physics}\ }\textbf {\bibinfo {volume} {89}},\ \bibinfo {pages} {7977--7982} (\bibinfo {year} {2001})}\BibitemShut {NoStop}%
\bibitem [{\citenamefont {Davydov}\ \emph {et~al.}(2002)\citenamefont {Davydov}, \citenamefont {Goncharuk}, \citenamefont {Smirnov}, \citenamefont {Nikolaev}, \citenamefont {Lundin}, \citenamefont {Usikov}, \citenamefont {Klochikhin}, \citenamefont {Aderhold}, \citenamefont {Graul}, \citenamefont {Semchinova} \emph {et~al.}}]{davydov2002composition}%
  \BibitemOpen
  \bibfield  {author} {\bibinfo {author} {\bibfnamefont {V.~Y.}\ \bibnamefont {Davydov}}, \bibinfo {author} {\bibfnamefont {I.}~\bibnamefont {Goncharuk}}, \bibinfo {author} {\bibfnamefont {A.}~\bibnamefont {Smirnov}}, \bibinfo {author} {\bibfnamefont {A.}~\bibnamefont {Nikolaev}}, \bibinfo {author} {\bibfnamefont {W.}~\bibnamefont {Lundin}}, \bibinfo {author} {\bibfnamefont {A.}~\bibnamefont {Usikov}}, \bibinfo {author} {\bibfnamefont {A.}~\bibnamefont {Klochikhin}}, \bibinfo {author} {\bibfnamefont {J.}~\bibnamefont {Aderhold}}, \bibinfo {author} {\bibfnamefont {J.}~\bibnamefont {Graul}}, \bibinfo {author} {\bibfnamefont {O.}~\bibnamefont {Semchinova}},  \emph {et~al.},\ }\bibfield  {title} {\enquote {\bibinfo {title} {Composition dependence of optical phonon energies and raman line broadening in hexagonal $al_xga_{1- x}n$ alloys},}\ }\href@noop {} {\bibfield  {journal} {\bibinfo  {journal} {Physical Review B}\ }\textbf {\bibinfo {volume} {65}},\ \bibinfo {pages} {125203} (\bibinfo {year}
  {2002})}\BibitemShut {NoStop}%
\bibitem [{\citenamefont {Ponce}, \citenamefont {Jena},\ and\ \citenamefont {Giustino}(2019)}]{ponce2019hole}%
  \BibitemOpen
  \bibfield  {author} {\bibinfo {author} {\bibfnamefont {S.}~\bibnamefont {Ponce}}, \bibinfo {author} {\bibfnamefont {D.}~\bibnamefont {Jena}}, \ and\ \bibinfo {author} {\bibfnamefont {F.}~\bibnamefont {Giustino}},\ }\bibfield  {title} {\enquote {\bibinfo {title} {Hole mobility of strained gan from first principles},}\ }\href@noop {} {\bibfield  {journal} {\bibinfo  {journal} {Physical Review B}\ }\textbf {\bibinfo {volume} {100}},\ \bibinfo {pages} {085204} (\bibinfo {year} {2019})}\BibitemShut {NoStop}%
\bibitem [{\citenamefont {Kyle}\ \emph {et~al.}(2014)\citenamefont {Kyle}, \citenamefont {Kaun}, \citenamefont {Burke}, \citenamefont {Wu}, \citenamefont {Wu},\ and\ \citenamefont {Speck}}]{kyle2014high}%
  \BibitemOpen
  \bibfield  {author} {\bibinfo {author} {\bibfnamefont {E.~C.}\ \bibnamefont {Kyle}}, \bibinfo {author} {\bibfnamefont {S.~W.}\ \bibnamefont {Kaun}}, \bibinfo {author} {\bibfnamefont {P.~G.}\ \bibnamefont {Burke}}, \bibinfo {author} {\bibfnamefont {F.}~\bibnamefont {Wu}}, \bibinfo {author} {\bibfnamefont {Y.-R.}\ \bibnamefont {Wu}}, \ and\ \bibinfo {author} {\bibfnamefont {J.~S.}\ \bibnamefont {Speck}},\ }\bibfield  {title} {\enquote {\bibinfo {title} {High-electron-mobility gan grown on free-standing gan templates by ammonia-based molecular beam epitaxy},}\ }\href@noop {} {\bibfield  {journal} {\bibinfo  {journal} {Journal of applied physics}\ }\textbf {\bibinfo {volume} {115}} (\bibinfo {year} {2014})}\BibitemShut {NoStop}%
\bibitem [{\citenamefont {Gotz}\ \emph {et~al.}(1998)\citenamefont {Gotz}, \citenamefont {Romano}, \citenamefont {Walker}, \citenamefont {Johnson},\ and\ \citenamefont {Molnar}}]{gotz1998hall}%
  \BibitemOpen
  \bibfield  {author} {\bibinfo {author} {\bibfnamefont {W.}~\bibnamefont {Gotz}}, \bibinfo {author} {\bibfnamefont {L.}~\bibnamefont {Romano}}, \bibinfo {author} {\bibfnamefont {J.}~\bibnamefont {Walker}}, \bibinfo {author} {\bibfnamefont {N.}~\bibnamefont {Johnson}}, \ and\ \bibinfo {author} {\bibfnamefont {R.}~\bibnamefont {Molnar}},\ }\bibfield  {title} {\enquote {\bibinfo {title} {Hall-effect analysis of gan films grown by hydride vapor phase epitaxy},}\ }\href@noop {} {\bibfield  {journal} {\bibinfo  {journal} {Applied physics letters}\ }\textbf {\bibinfo {volume} {72}},\ \bibinfo {pages} {1214--1216} (\bibinfo {year} {1998})}\BibitemShut {NoStop}%
\bibitem [{\citenamefont {Jhalani}\ \emph {et~al.}(2020)\citenamefont {Jhalani}, \citenamefont {Zhou}, \citenamefont {Park}, \citenamefont {Dreyer},\ and\ \citenamefont {Bernardi}}]{jhalani2020piezoelectric}%
  \BibitemOpen
  \bibfield  {author} {\bibinfo {author} {\bibfnamefont {V.~A.}\ \bibnamefont {Jhalani}}, \bibinfo {author} {\bibfnamefont {J.-J.}\ \bibnamefont {Zhou}}, \bibinfo {author} {\bibfnamefont {J.}~\bibnamefont {Park}}, \bibinfo {author} {\bibfnamefont {C.~E.}\ \bibnamefont {Dreyer}}, \ and\ \bibinfo {author} {\bibfnamefont {M.}~\bibnamefont {Bernardi}},\ }\bibfield  {title} {\enquote {\bibinfo {title} {Piezoelectric electron-phonon interaction from ab initio dynamical quadrupoles: Impact on charge transport in wurtzite gan},}\ }\href@noop {} {\bibfield  {journal} {\bibinfo  {journal} {Physical Review Letters}\ }\textbf {\bibinfo {volume} {125}},\ \bibinfo {pages} {136602} (\bibinfo {year} {2020})}\BibitemShut {NoStop}%
\bibitem [{\citenamefont {Jena}\ \emph {et~al.}(2003)\citenamefont {Jena}, \citenamefont {Heikman}, \citenamefont {Speck}, \citenamefont {Gossard}, \citenamefont {Mishra}, \citenamefont {Link},\ and\ \citenamefont {Ambacher}}]{jena2003magnetotransport}%
  \BibitemOpen
  \bibfield  {author} {\bibinfo {author} {\bibfnamefont {D.}~\bibnamefont {Jena}}, \bibinfo {author} {\bibfnamefont {S.}~\bibnamefont {Heikman}}, \bibinfo {author} {\bibfnamefont {J.~S.}\ \bibnamefont {Speck}}, \bibinfo {author} {\bibfnamefont {A.}~\bibnamefont {Gossard}}, \bibinfo {author} {\bibfnamefont {U.~K.}\ \bibnamefont {Mishra}}, \bibinfo {author} {\bibfnamefont {A.}~\bibnamefont {Link}}, \ and\ \bibinfo {author} {\bibfnamefont {O.}~\bibnamefont {Ambacher}},\ }\bibfield  {title} {\enquote {\bibinfo {title} {Magnetotransport properties of a polarization-doped three-dimensional electron slab in graded {Al}{Ga}{N}},}\ }\href@noop {} {\bibfield  {journal} {\bibinfo  {journal} {Physical Review B}\ }\textbf {\bibinfo {volume} {67}},\ \bibinfo {pages} {153306} (\bibinfo {year} {2003})}\BibitemShut {NoStop}%
\bibitem [{\citenamefont {Simon}\ \emph {et~al.}(2006)\citenamefont {Simon}, \citenamefont {Wang}, \citenamefont {Xing}, \citenamefont {Rajan},\ and\ \citenamefont {Jena}}]{simon2006carrier}%
  \BibitemOpen
  \bibfield  {author} {\bibinfo {author} {\bibfnamefont {J.}~\bibnamefont {Simon}}, \bibinfo {author} {\bibfnamefont {A.~K.}\ \bibnamefont {Wang}}, \bibinfo {author} {\bibfnamefont {H.}~\bibnamefont {Xing}}, \bibinfo {author} {\bibfnamefont {S.}~\bibnamefont {Rajan}}, \ and\ \bibinfo {author} {\bibfnamefont {D.}~\bibnamefont {Jena}},\ }\bibfield  {title} {\enquote {\bibinfo {title} {Carrier transport and confinement in polarization-induced three-dimensional electron slabs: Importance of alloy scattering in {Al}{Ga}{N}},}\ }\href@noop {} {\bibfield  {journal} {\bibinfo  {journal} {Applied physics letters}\ }\textbf {\bibinfo {volume} {88}} (\bibinfo {year} {2006})}\BibitemShut {NoStop}%
\bibitem [{\citenamefont {Kaneki}\ \emph {et~al.}(2024)\citenamefont {Kaneki}, \citenamefont {Konno}, \citenamefont {Kimura}, \citenamefont {Kanegae}, \citenamefont {Suda},\ and\ \citenamefont {Fujikura}}]{kaneki2024record}%
  \BibitemOpen
  \bibfield  {author} {\bibinfo {author} {\bibfnamefont {S.}~\bibnamefont {Kaneki}}, \bibinfo {author} {\bibfnamefont {T.}~\bibnamefont {Konno}}, \bibinfo {author} {\bibfnamefont {T.}~\bibnamefont {Kimura}}, \bibinfo {author} {\bibfnamefont {K.}~\bibnamefont {Kanegae}}, \bibinfo {author} {\bibfnamefont {J.}~\bibnamefont {Suda}}, \ and\ \bibinfo {author} {\bibfnamefont {H.}~\bibnamefont {Fujikura}},\ }\bibfield  {title} {\enquote {\bibinfo {title} {Record high electron mobilities in high-purity gan by eliminating c-induced mobility collapse},}\ }\href@noop {} {\bibfield  {journal} {\bibinfo  {journal} {Applied Physics Letters}\ }\textbf {\bibinfo {volume} {124}} (\bibinfo {year} {2024})}\BibitemShut {NoStop}%
\bibitem [{\citenamefont {Bagheri}\ \emph {et~al.}(2022)\citenamefont {Bagheri}, \citenamefont {Qui{\~n}ones-Garcia}, \citenamefont {Khachariya}, \citenamefont {Rathkanthiwar}, \citenamefont {Reddy}, \citenamefont {Kirste}, \citenamefont {Mita}, \citenamefont {Tweedie}, \citenamefont {Collazo},\ and\ \citenamefont {Sitar}}]{bagheri2022high}%
  \BibitemOpen
  \bibfield  {author} {\bibinfo {author} {\bibfnamefont {P.}~\bibnamefont {Bagheri}}, \bibinfo {author} {\bibfnamefont {C.}~\bibnamefont {Qui{\~n}ones-Garcia}}, \bibinfo {author} {\bibfnamefont {D.}~\bibnamefont {Khachariya}}, \bibinfo {author} {\bibfnamefont {S.}~\bibnamefont {Rathkanthiwar}}, \bibinfo {author} {\bibfnamefont {P.}~\bibnamefont {Reddy}}, \bibinfo {author} {\bibfnamefont {R.}~\bibnamefont {Kirste}}, \bibinfo {author} {\bibfnamefont {S.}~\bibnamefont {Mita}}, \bibinfo {author} {\bibfnamefont {J.}~\bibnamefont {Tweedie}}, \bibinfo {author} {\bibfnamefont {R.}~\bibnamefont {Collazo}}, \ and\ \bibinfo {author} {\bibfnamefont {Z.}~\bibnamefont {Sitar}},\ }\bibfield  {title} {\enquote {\bibinfo {title} {High electron mobility in aln: Si by point and extended defect management},}\ }\href@noop {} {\bibfield  {journal} {\bibinfo  {journal} {Journal of Applied Physics}\ }\textbf {\bibinfo {volume} {132}} (\bibinfo {year} {2022})}\BibitemShut {NoStop}%
\bibitem [{\citenamefont {Taniyasu}, \citenamefont {Kasu},\ and\ \citenamefont {Makimoto}(2006)}]{taniyasu2006increased}%
  \BibitemOpen
  \bibfield  {author} {\bibinfo {author} {\bibfnamefont {Y.}~\bibnamefont {Taniyasu}}, \bibinfo {author} {\bibfnamefont {M.}~\bibnamefont {Kasu}}, \ and\ \bibinfo {author} {\bibfnamefont {T.}~\bibnamefont {Makimoto}},\ }\bibfield  {title} {\enquote {\bibinfo {title} {Increased electron mobility in n-type si-doped aln by reducing dislocation density},}\ }\href@noop {} {\bibfield  {journal} {\bibinfo  {journal} {Applied physics letters}\ }\textbf {\bibinfo {volume} {89}} (\bibinfo {year} {2006})}\BibitemShut {NoStop}%
\end{thebibliography}%

\end{document}